\documentclass[11pt]{article}
\pdfoutput=1
\usepackage{aas_macros,amsmath,amssymb,comment,cite,esint,graphicx,mathtools}
\usepackage[margin=.8in,letterpaper]{geometry}
\usepackage[colorlinks=true]{hyperref}
\usepackage[affil-it]{authblk}
\usepackage{subcaption}
\usepackage[utf8]{inputenc}
\usepackage{mathrsfs}
\usepackage{appendix}
\usepackage{amssymb}
\usepackage{float}
\usepackage{subcaption}
\usepackage{xcolor}
\usepackage{cite}
\usepackage{hyperref}
\hypersetup{pageanchor=false}
\usepackage{indentfirst}
\usepackage{url}
\usepackage{float}
\usepackage{caption}
\usepackage[numbers,square,comma,sort&compress,merge]{natbib}
\usepackage{esint}
\usepackage{overpic}
\usepackage{graphicx}
\usepackage{epsf,amsmath,bbold,amsfonts,stmaryrd}
\usepackage{textcomp}
\usepackage{ulem}
\usepackage{tikz}
\usepackage{amsfonts}
\usepackage{array}
\usepackage{longtable}
\usepackage[utf8]{inputenc}
\usepackage{multirow}
\usepackage{orcidlink}
\numberwithin{equation}{section}
\let\originalleft\left
\let\originalright\right
\renewcommand{\left}{\mathopen{}\mathclose\bgroup\originalleft}
\renewcommand{\right}{\aftergroup\egroup\originalright}
\mathcode`\*="8000
{\catcode`\*=\active\gdef*{\mathclose{}\,\mathopen{}}}

\newcommand{\be}{\begin{equation}}
\newcommand{\ee}{\end{equation}}
\newcommand{\bea}{\setlength\arraycolsep{2pt} \begin{eqnarray}}
\newcommand{\eea}{\end{eqnarray}}
\newcommand{\nn}{\nonumber}

\def\l{\left}

\def\nn{\nonumber}

\def\r{\right}

\def\be{\begin{equation}}
\def\ee{\end{equation}}
\def\bag{\begin{aligned}}
\def\eag{\end{aligned}}
\def\bea{\begin{eqnarray}}
\def\eea{\end{eqnarray}}
\def\ba{\begin{array}}
\def\ea{\end{array}}

\def\bc{\begin{center}}
\def\ec{\end{center}}

\begin{document}
\title{Probing Observable Features of Lorentz violation in Low-Energy Ho\v{r}ava Gravity with Accretion Disk Images of Black Hole}

\author{Meng-Die Zhao$^{1\dag}$,
Yu-Yan Wang$^{1\ddag}$, 
Ke-Jian He$^{2\ast}$and
Guo-Ping Li$^{1\S}$}
\date{}

\maketitle

\vspace{-10mm}

\begin{center}
{\textit
\textsuperscript{$^1$}School of Physics and Astronomy, China West Normal University, Nanchong 637000, China\\\vspace{4mm}
\textsuperscript{2}Department of Mechanics, Chongqing Jiaotong University, Chongqing 400074, China}
\end{center}

\vspace{8mm}

\begin{abstract}

In this paper, we study the observable signatures of Lorentz violation (LV) in low-energy Ho\v{r}ava gravity by simulating the images and polarization features of rotating LV black holes using a backward ray-tracing method. Within a thin-disk accretion model and the ZAMO framework, we numerically solve the geodesics equation of photon and simulated the corresponding thin-disk images and polarization patterns. The results show that the LV parameter $\ell$ strongly affects the inner shadow, brightness asymmetry, and polarization properties of the thin disk. The decrease of $\ell$ leads to a more elliptical and untilted inner shadow, while increasing $\ell$ produces a pronounced leftward ``D''-shaped structure of critical curve. And, the variation of $\ell$ alters the distribution of polarized intensity and polarization direction, especially near the critical curve. More importantly, it also shows that a positive $\ell$ enhances the black hole’s angular velocity, while a negative one suppresses it, indicating that the sign of $\ell$ determines the trend direction of the LV effect. 
These findings suggest that future high-resolution EHT observations combining the thin-disk images and polarization patterns could provide valuable tests of the LV effect.
\end{abstract}

\vfill
\begin{flushleft} 
    \footnotesize
    $\dag$ e-mail: \href{mailto:zhaomengdie0926@163.com}{zhaomengdie0926@163.com} \\
    $\ddag$ e-mail: \href{mailto:wangyy2027@163.com}{wangyy2027@163.com} \\
    $\ast$ e-mail: \href{mailto:kjhe94@163.com}{kjhe94@163.com}\\
     $\S$ e-mail: \href{mailto:gpliphys@yeah.net}{gpliphys@yeah.net} (corresponding author)
\end{flushleft}

\maketitle

\newpage
\baselineskip 18pt
\section{Introduction}\label{sec1}
Since Einstein’s formulation of general relativity (GR), the study of gravitational waves and black holes has evolved into a key avenue for exploring strong gravitational fields and the structure of spacetime, driving continuous theoretical and observational advances in our understanding of the universe.
As a landmark achievement, LIGO’s first detection of GW150914 \cite{LIGOScientific:2016aoc}, produced by a binary black hole merger, not only allowed us to hear the so-called ``sound” of black holes but also opened the era of gravitational-wave astronomy.
As compelling visual evidence for the existence of black holes, the Event Horizon Telescope (EHT) collaboration released the first image of the black hole shadow at the center of the galaxy M87* in the Virgo cluster in 2019\cite{EventHorizonTelescope:2019ths,EventHorizonTelescope:2019dse}, thereby revealing, for the first time, the optical observational features of a black hole. Three years later, the EHT collaboration released an image of Sagittarius A* (Sgr A*)\cite{EventHorizonTelescope:2022xqj}, further confirming the imaging techniques previously applied to M87*, while also highlighting the more dynamic and variable characteristics of Sgr A*. The visual depiction of these black holes not only confirmed the predictions of GR but also provided new opportunities for investigating the distribution of matter and electromagnetic phenomena in the vicinity of black holes. 

In both sets of black hole images released by EHT, a prominent bright ring structure, known as the bright ring, consistently appears, enclosing a central dark region referred to as the black hole shadow. The shadow, as an intrinsic observable feature of black holes, has long been a central focus of both theoretical and observational investigations.
In 1966, Synge analyzed the shadow cast by a Schwarzschild black hole and derived an expression for its angular radius\cite{Synge:1966okc}. Subsequently, Bardeen extended this analysis to Kerr black holes, demonstrating that their shadows deviate from perfect circularity\cite{Bardeen:1973tla}. Further developments were made by Hioki and Maeda, who examined the shadows of Kerr black holes and naked singularities\cite{hioki2009measurement}.
Motivated by this progress, one has investigated the shadow size and shape of a wide range of black hole models, spanning black holes in different gravity theories, those coupled to some matter fields, regular black holes, and so on\cite{Johannsen:2010ru,Haroon:2018ryd,Dymnikova:2019vuz,Kumar:2019pjp,Kumar:2020owy,Hioki:2023ozd}. Such analyses provide systematic comparisons of their deviations from the Kerr spacetime and explore the constraints that observational data may impose on the model parameters\cite{EventHorizonTelescope:2021dqv,Afrin:2021wlj,Kuang:2022xjp}. These findings highlight the essential importance of black hole shadows as a powerful tool for uncovering the nature of gravity and revealing the key features of black hole spacetimes.
In addition, for the bright ring reported by the EHT, there exists an extremely thin luminous substructure corresponding to the black hole’s critical curve, commonly referred to as the photon ring. 
Because the formation of photon ring is determined solely by the spacetime, its properties have become a major focus of recent research, leading to a number of significant developments\cite{Peng:2020wun,Wielgus:2021peu,Gan:2021xdl,Peng:2021osd,Zeng:2024ptv,Detournay:2025xqd}. In particular, the study in 2022 suggested that the photon ring may exhibit a form of holographic behavior\cite{Hadar:2022xag}, implying that it could encode deeper physical information within the frameworks of holography. It is clear that, the photon ring is emerging as one of the most promising interfaces between theoretical and observational studies in strong-gravity physics.
In general, black holes do not emit light, and their observable appearance must therefore be revealed through the radiation from the surrounding accreting material. In theoretical studies, one has proposed several mainstream accretion-flow models to describe the matter distribution near black holes, including spherical accretion\cite{Zeng:2020dco,Gan:2021pwu,Meng:2023htc}, thin-disk models\cite{Pun:2008ae,Gao:2023mjb,Liu:2021yev,Zeng:2021dlj,Uniyal:2022vdu,Heydari-Fard:2022xhr,Yang:2024nin,Guo:2024mij}, thick-disk models\cite{Zhang:2024jrw,Zhang:2024lsf}, and jet models\cite{Hou:2023bep,Zhang:2024lsf}, among others. 
For the study of thin disks, particularly in 2019, Wald et al. shown that the bright ring outside the black hole shadow is composed of direct emission, the lensing ring, and the photon ring\cite{Gralla:2019xty}. 
They carefully analyzed the luminosity accumulation effect arising from light rays traversing the thin disk and further characterized the disk image of a Schwarzschild black hole.
Subsequently, one further extended this work to the spacetimes in modified gravity theories, and those coupled with matter fields\cite{Liu:2020vkh,Gyulchev:2021dvt,li2021observational,Hou:2022eev,Zeng:2023fqy,Heydari-Fard:2023kgf,Liu:2025hhg}, as well as the rotating black holes\cite{Hou:2022eev,Liu:2025hhg}, further encompassing various compact objects such as boson stars, wormholes, and so on\cite{Rosa:2023qcv,Guerrero:2022qkh,Rosa:2022tfv}. 
These advancements have provided new theoretical avenues for interpreting and clarifying observations related to black hole images.



On the other hand, the linear-polarimetric images of M87* reconstructed from the EHT 1.3 mm observations in Ref. \cite{EventHorizonTelescope:2021bee} show that only a portion of the bright ring exhibits significant polarization and undergoes temporal evolution, thereby revealing key information about the magnetic field structure within the near-horizon relativistic plasma.
The study of black hole polarization images not only helps to reveal the magnetic field structures in their vicinity, but is also closely related to the properties of the accreting matter and the strong gravitational field itself\cite{Zhu:2022amy,Hu:2022sej,Chael:2023pwp,Zhang:2021hit}.
Based on a synchrotron radiation model of Kerr black hole immersed in a uniform magnetic field, Chen et al.\cite{Zhang:2023cuw} simulates the spontaneously vortical motions of charged particles in the jet region and reconstructs their polarized images, thereby providing a new physical mechanism for understanding the electromagnetic processes in jets and their contribution to black hole imaging.
And, for a rotating black hole surrounded by a cold dark matter halo, Ref.\cite{Qin:2023nog} simulates its polarization signatures of an equatorial emitting ring  revealing that black hole polarization images serve as a potential—but not highly sensitive—probe of dark matter effects.
Besides this, the polarized images of other compact objects have also been further investigated. Notably, we studied the polarized images of boson stars, whose distinctive polarization features may provide a possible means to distinguish them from black holes\cite{Li:2025awg}.
Taken together, these works demonstrate that polarization observations can not only unveil the electromagnetic and dynamical structures surrounding black holes but also serve as an effective probe for testing gravity and distinguishing compact object models\cite{EventHorizonTelescope:2021btj,Gelles:2021kti,Yang:2025byw}.
In addition, the Lorentz Violation(LV), as a significant phenomenon within the framework of quantum physics, the theoretical research and discussions of it increasing, and various models have been introduced in the process\cite{Kostelecky:2003fs,PhysRevD.58.116002,PhysRevD.97.104001,PhysRevD.103.044002,PhysRevD.108.124004,Duan:2023gng,Liu_2024,Ara_jo_Filho_2024,Liu:2025fxj}. 
For example, spherically symmetric black hole solutions in both Bumblebee and Kalb–Ramond gravity have been obtained, thereby enriching our understanding of black hole physics in LV frameworks\cite{PhysRevD.97.104001,PhysRevD.103.044002,PhysRevD.108.124004,Duan:2023gng,Liu_2024}.
And, the effect of LV have been discussed widely, such as gravitational lensing\cite{Junior:2024vdk,Mangut:2023oxa}, black hole shadows\cite{Liu:2024lve,Liu:2025lwj}, Hawking radiation \cite{Baruah:2025ifh}, and quasinormal modes\cite{Oliveira:2021abg,Baruah:2023rhd}. 
Analyzing these deviations not only helps to constrain the magnitude of LV but also highlights its essential role in both theoretical studies and observational analyses.
Last year, Devecio\v{g}lu and Park employed a concise two-step method to derive the exact Kerr-type rotating black hole solution in the low-energy limit of Ho\v{r}ava gravity\cite{Devecioglu:2024uyi}, which remains asymptotically flat at infinity.  
Since black hole shadow observations provide one of the most direct probes of strong gravitational fields, the shadow of this LV rotating black hole has been studied under celestial illumination using numerical ray tracing. Preliminary constraints on the LV parameters have been obtained by comparing the results with the EHT observations of M87* and Sgr A*\cite{Liu:2025lwj}, highlighting the importance of investigating its observable signatures.

In fact, both black hole shadow and inner shadow arise from strong gravitational lensing, with their boundaries determined by the structure of photon orbits. However, the former corresponds to the entire dark region enclosed by the critical curve in spherical emission models, whereas the latter appears in thin-disk represents the direct lensed image of the equatorial portion of the event horizon\cite{Chael:2021rjo}. The inner shadow is smaller in size and more closely tied to the actual horizon geometry, thereby capturing deeper characteristic features of the black hole. 
Therefore, a further investigation of the inner shadow of LV rotating black holes \cite{Devecioglu:2024uyi} is particularly significant for revealing the nature of LV effects. On the other hand, since polarized images of black hole encodes not only the magnetic field structure but also intrinsic properties of the spacetime geometry\cite{Zeng:2025kyv}, an in-depth study of the polarized images of such black hole is equally valuable. Motivated by these considerations, in this work we employ a backward ray-tracing approach within the framework of a thin accretion disk model to systematically examine both the image and polarization features of LV rotating black hole. Our aim is to extract key observables associated with LV effects—including the inner shadow, photon ring, and polarization patterns—and use these quantities to conduct a comprehensive analysis of the LV signatures. It is anticipated that our results will provide a foundation for testing LV effects with future high-resolution astronomical observations.


The remainder of this paper is organized as follows. In Section \ref{sec2}, we review the properties of rotating LV black holes. Section \ref{sec3} introduces the black hole imaging methods employed in this study. In Section \ref{sec4}, we describe the imaging techniques for the thin accretion disk and analyze its intensity distribution and redshift characteristics. Section \ref{sec5} focuses on the generation of polarization images, providing a detailed analysis of the polarization intensity and orientation. In Section \ref{sec6}, we present a comparative study between rotating LV black holes and Kerr black holes, emphasizing their respective images and intensity distributions. Finally, Section \ref{sec7} summarizes our main findings.

\section{ The rotating Lorentz-Violating black hole }\label{sec2}
The action that characterizes the low-energy sector of non-projectable Ho\v{r}ava gravity, modulo boundary contributions, is expressed as\cite{PhysRevD.79.084008}
\bea\label{l1}
S_g = \int_{\mathbb{R} \times \Sigma_t} dtd^3x \, \sqrt{g} N \left[ \frac{1}{\kappa} \left( K_{ij} K^{ij} - \lambda K^2 \right) + \xi R \right],
\eea
with $K_{ij} = (2N)^{-1}(\dot{g}_{ij} - \nabla_i N_j - \nabla_j N_i)$ denotes the extrinsic curvature, $\xi$, $\lambda$ and $\kappa$ are the coupling constants, and $R$ represents the three-dimensional Ricci scalar in the ADM metric, which reads
\bea\label{andl2}
ds^2 = -N^2 dt^2 + g_{ij} \left( dx^i + N^i dt \right) \left( dx^j + N^j dt \right).
\eea
The equations from the variations of \( N \), \( N^i \), and \( g^{ij} \) are given by
\begin{align}\label{andl3}
& H \equiv \frac{1}{\kappa} \left( K_{ij} K^{ij} - \lambda K^2 \right) - \xi R = 0, \\
& H^i \equiv -\frac{2}{\kappa} \nabla_j \left( K^{ji} - \lambda K g^{ji} \right) = 0, \\
&E_{ij} \equiv \frac{1}{\kappa} \left( E^{(1)}_{ij} - \lambda E^{(2)}_{ij} \right) + \xi E^{(3)}_{ij}= 0,
\end{align}
with 
\begin{align}\label{andl4}
E_{ij}^{(1)} &= N_i \nabla_k K^k_j + N_j \nabla_k K^k_i - K^k_i \nabla_j N_k 
 - K^k_j \nabla_i N_k - N^k \nabla_k K_{ij} \nonumber \\
&- 2N K_{ik} K_j^k - \frac{1}{2} N K^{k\ell} K_{\ell k} g_{ij} + N K K_{ij} + \dot{K}_{ij}, \\
E_{ij}^{(2)} &= \frac{1}{2} N K^2 g_{ij} + N_i \partial_j K + N_j \partial_i K - N^k (\partial_k K) g_{ij} + \dot{K}_{ij}, \\
E_{ij}^{(3)} &= N \left( R_{ij} - \frac{1}{2} R g_{ij}\right) - (\nabla_i \nabla_j - g_{ij} \nabla_k \nabla^k N) N, \\
E_{ij}^{(4)} &= \frac{1}{N} \left( -\frac{1}{2} g_{ij} \nabla_k N \nabla^k N + \nabla_i N \nabla_j N \right).
\end{align}
The construction of rotating black hole solutions begins with the massless case. Notably, it has recently been shown that the massless Kerr solution is also a valid solution of Ho\v{r}ava gravity, irrespective of the value of $\lambda$ \cite{Devecioglu:2024uyi}. We have 
\begin{align}\label{andl5}
ds_0^2 = -\frac{\rho^2 \Delta_r^{(0)}}{\Sigma_{(0)}^2} dt^2 + \frac{\rho^2}{\Delta_r^{(0)}} dr^2 + \rho^2 d\theta^2 + \frac{\Sigma_{(0)}^2 \sin^2 \theta}{\rho^2} d\phi^2
\end{align}
where \( \rho^2 = r^2 + a^2 \cos^2\theta, \, \Delta_r^{(0)} = (r^2 + a^2), \, \Sigma_{(0)}^{2} = (r^2 + a^2)\rho^2 \), and \( (t, r, \theta, \phi) \) are the Boyer-Lindquist coordinates. Eq.(\ref{andl5}) is essentially Minkowski spacetime written in ellipsoidal coordinates with parameter $a$. These coordinates extend to $r < 0$ and feature another asymptotic infinity, so the massless Kerr metric has been interpreted as a wormhole-like geometry.

The next step is to adopt a more general metric ansatz by introducing mass-dependent functions, so as to obtain a consistent rotating black-hole solution. Since in the Kerr metric the mass term is tightly intertwined with the spin parameter $a$ and cannot be cleanly disentangled, it is necessary to employ a generic setup that clearly separates the spin and mass sectors, which might crack the rigid structure of the Kerr solution. Following the Kerr solution, we introduce the metric ansatz
\begin{align}\label{andl6}
ds_1^2 = -N^2 dt^2 + \frac{\rho^2}{\Delta_r} dr^2 + \rho^2 d\theta^2 + \frac{\Sigma^2 \sin^2 \theta}{\rho^2} \left( d\phi + N^\phi dt \right)^2,
\end{align}
with
\begin{align}\label{andl7}
\Sigma^2 = \left( r^2 + a^2 \right) \rho^2 + f(r) a^2 \sin^2\theta, \quad  N^2 = \frac{\rho^2 \Delta_r(r)}{\Sigma^2}, \quad N^\phi = -\frac{g(r)}{\Sigma^2}.
\end{align}
Based on Eqs.(\ref{andl6}) and (\ref{andl7}), we can solve the full equations of motion (\ref{andl3}) and obtain
\begin{align}\label{andl8}
f(r) = 2Mr, \quad g(r) = 2aMr\sqrt{\kappa\xi}, \quad \Delta_r(r) = r^2 + a^2 - 2Mr,
\end{align}
here $M$ is an integration constant and can be regarded as the mass
parameter of the black hole. And when $\xi = 1/\kappa$, the solution reduces to the Kerr solution of general relativity; when $a = 0$, it will further reduce to the Schwarzschild solution.
Here, we adopt the standard asymptotic normalization conditions
$N^\varphi\big|_{\infty} = 0, W(\infty) \equiv N\sqrt{g_{rr}}\big|_{\infty} = 1.$
It is worth noting that all corrections involving the Lorentz-violating factor $\sqrt{\kappa \xi}$ are entirely encoded in the shift function $N^\varphi$. Then, the metric can be recast into the following form,
\begin{equation}\label{andl9}
\begin{aligned}
ds_1^2 ={}&
\left[
- \frac{\Delta - a^2 \sin^2\theta}{\rho^2}
+ \frac{(\kappa \xi - 1)(2Mr)^2 a^2 \sin^2\theta}{\rho^2 \Sigma^2}
\right] dt^2
+ \frac{\rho^2}{\Delta} dr^2
+ \rho^2 d\theta^2  \\
&+ \frac{\Sigma^2 \sin^2\theta}{\rho^2} d\phi^2
- \frac{4 a M r \sqrt{\kappa \xi} \sin^2\theta}{\rho^2} \, dt\, d\phi .
\end{aligned}
\end{equation}

It is noteworthy that as long as $\xi \neq 1/\kappa$, the $g_{tt}$ and $g_{t\phi}$ components exhibit a nontrivial Lorentz-symmetry–violating effect. 
For convenience in the following discussion, we define ${{\ell} \equiv \sqrt{\kappa \xi}} - 1$, thereby the metric form in Eq.\ref{andl9} can be taken a expression with inclusion of the Kerr metric as follows,
\bea
d s_2^2 = d s_{\mathrm{kerr}}^2
+ \frac{4 \ell a M r \sin^2\theta}{\rho^2}
\left(
\frac{(\ell+2)\,a M r}{\Sigma}\,dt^2 - dt\,d\varphi
\right),
\label{2.5}
\eea
where $ds^2_{\text{kerr}}$ represents the metric of the Kerr black hole. 
In this paper, the parameter $\ell$ characterizes a nontrivial Lorentz-symmetry--violating effect in spacetime and is referred to as the Lorentz-Violating(LV) parameter. When $l = 0$ (i.e., $\xi = 1/\kappa$), the correction terms introduced by Lorentz violation vanish completely, and the spacetime reduces to the Kerr black hole. It is worth noting that, in Boyer--Lindquist coordinates, this black hole possesses two horizons, $r_\pm$, whose locations are identical to those of the Kerr black hole. The horizons are determined by the equation $\Delta = 0$, yielding
$r_\pm = M \pm \sqrt{M^2 - a^2}$. Previous studies have shown that, although the horizon locations remain unchanged, the shadow shape of the black hole exhibits significant variation as the LV parameter $\ell$ increases\cite{Liu:2025lwj}. Motivated by this result, our aim is to further investigate whether the inner shadow and observable images of this black hole differ markedly from those of the Kerr black hole in the presence of a thin-disk background. For simplicity, the black hole mass is set to $M = 1$ in the subsequent analysis.

In this work, we focus on probing the observable features of Lorentz symmetry violation in low-energy Ho\v{r}ava gravity. Specifically, to obtain the black hole image, we first need to solve the propagation behavior of light. And, the propagation of light can be described by null geodesics. The corresponding constrained Hamiltonian for the geodesics can be expressed in the following form,
\bea
\mathcal{H} = \frac{1}{2} g^{\mu\nu} p_{\mu} p_{\nu},
\eea
where $g_{\mu\nu}$ is the metric corresponding to the line element in Eq.\ref{2.5}, and $p_{\mu}$ is the dual of the photon’s four-momentum vector, defined as $p_{\mu} \equiv g_{\mu\nu} p^{\nu}$. 
The photon’s four-momentum vector reads $p^{\mu} = (\dot{t}, \dot{r}, \dot{\theta}. \dot{\varphi})$. The symbol “\(\cdot\)” denotes the derivative with respect to the affine parameter.
And, the photon Hamiltonian constraint condition is
\begin{align}
\mathcal{H} = {} &
\frac{\left(\mathcal{L}\csc\theta - a\mathcal{E}\sin\theta\right)^{2}}{2\rho^{2}}
- \frac{\left[(r^{2}+a^{2})\mathcal{E} - a\mathcal{L}\right]^{2}}{2\rho^{2}\Delta}
+ \frac{\rho^{2}\dot{r}^{2}}{2\Delta}
+ \frac{\rho^{2}\dot{\theta}^{2}}{2}
+ \frac{2\mathcal{L}Mra \ell \mathcal{E}}{\rho^{2}\Delta} \nonumber \\[6pt]
& - \frac{2\mathcal{L}^{2}M^{2}r^{2}a^{2} \ell (\ell+2)}{\rho^{2}\Delta\Sigma^{2}}
= 0.
\end{align}
We note that this spacetime admits two Killing vector fields, namely, 
$\xi^{\mu} = (\partial / \partial t)^{\mu}$ and 
$\psi^{\mu} = (\partial / \partial \phi)^{\mu}$, 
which correspond to two conserved quantities, $\mathcal{E}$ and $\mathcal{L}$, respectively. 
They are given by
$
\mathcal{E} = -g_{\mu\nu}\xi^\mu\dot{x}^\nu, \mathcal{L} = g_{\mu\nu}\psi^\mu\dot{x}^\nu.
$
Since we adopt the backward ray-tracing method to image the black hole, it is necessary to perform a backward integration of the equations of motion for photons along the null geodesics starting from the observer. This is equivalent to solving the photon trajectories by backward integration of the Hamiltonian equations, which are
\bea
\frac{\partial \mathcal{H}}{\partial p_\mu} = \dot{x}^\mu, \quad \frac{\partial \mathcal{H}}{\partial x^\mu} = -\dot{p}_\mu.
\eea
When performing the backward integration of the above equations of motion, it is necessary to set up a camera model at the observer  
that maps the photon momentum \(p_\mu\) onto the screen via stereographic projection.  
Once the corresponding pixel positions on the screen are determined, the evolution of each light ray can be traced.  
A detailed description of this imaging method will be provided in the next.

\section{Black Hole Imaging Framework: Camera Model and Field of View Configuration}\label{sec3}
\begin{figure}[h]
    \centering
    \begin{subfigure}[b]{0.45\textwidth}
    \includegraphics[width=\textwidth]{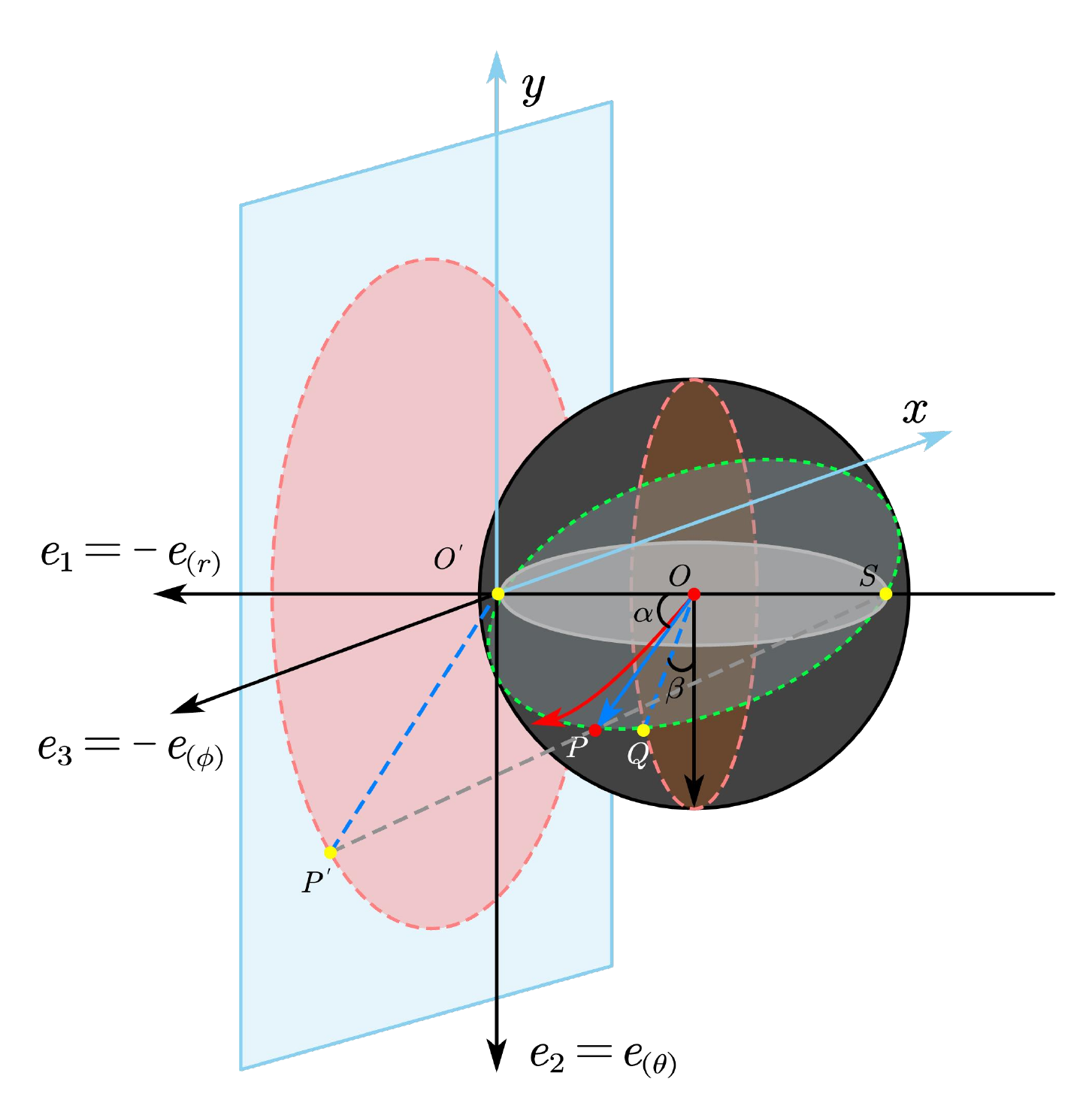}
    \end{subfigure}
    \caption{The method of stereographic projection}
    \label{fig:1}
\end{figure}

In this section, we provide a detailed exposition of black hole imaging techniques, with a particular focus on the construction of the camera model and the configuration of pixel positions.
First, we consider a zero angular momentum observer (ZAMO) situated at a radial coordinate $r = r_o$. The observer’s position is specified by the coordinates  $({t}_{o},{r}_{o}, {\theta }_{o}, {\phi }_{o})$, with ${t}_{o}=0$ and ${\phi }_{o}=0$. 
Then, we construct a local orthonormal reference frame at observer’s position, with the basis vector set denoted by $\{{ e_{0}, e_{1}, e_{2}, e_{3} }\}$. 
This reference frame can be mapped to the coordinate basis of the black hole spacetime $e_{{({\mu})}}({\partial_{0}, \partial_{1}, \partial_{2}, \partial_{3} })$ via the transformation relation $e_{({\mu})} = e_{({\mu})}^{ \nu} \partial_{\nu}$, where the transformation matrix $e_{{({\mu})}}^{\nu}$ satisfies the following relation
\bea
g_{\mu \nu} e_{({\alpha})}^{{\mu}} e_{({\beta})}^{\nu} = \eta_{({\alpha}) ({\beta})},
\eea
where $\eta_{{(\alpha)} {(\beta)}}$ is the Minkowski spacetime metric. And, the explicit form is given by
\bea\label{e32}
{e_{0}}= \frac{g_{\phi\phi}\partial_{t}-g_{t\phi}\partial_{\phi}}{\sqrt{g_{\phi\phi}(g_{t\phi}^{2}-g_{\phi\phi}g_{tt})}}, \quad
e_{1}=-\frac{\partial_{r}}{\sqrt{g_{rr}}},\quad
e_{2}= \frac{\partial_{\theta}}{\sqrt{g_{\theta\theta}}},\quad
e_{3}= - \frac{\partial_{\phi}}{\sqrt{g_{\phi\phi}}}.
\eea
According to \cite{Li:2024ctu}, as shown in Fig.\ref{fig:1}, when a light ray reaches the point $O$, its momentum is given by $\overrightarrow{OP}$. 
We construct a unit sphere with $\overrightarrow{OP}$ as the radius and $O$ as the center, 
and set up an orthonormal basis at $O'$ on the corresponding screen (as in Eq.\ref{e32}). 
The celestial coordinates $\alpha$ and $\beta$ are defined as the angles between the light ray direction and the basis vectors. 
The construction of the unit sphere, orthonormal basis as well as celestial coordinates are illustrated in the following figure, which is 

\begin{figure}[h]
    \centering
    \begin{subfigure}[b]{0.45\textwidth}
        \includegraphics[width=\textwidth]{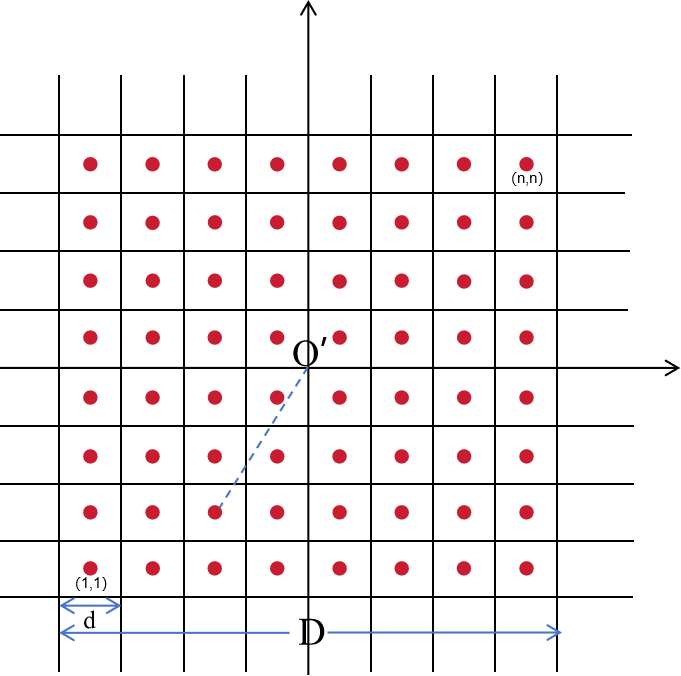}
        \caption{Cartesian plane}
    \end{subfigure}
    \quad
    \begin{subfigure}[b]{0.45\textwidth}
        \includegraphics[width=\textwidth]{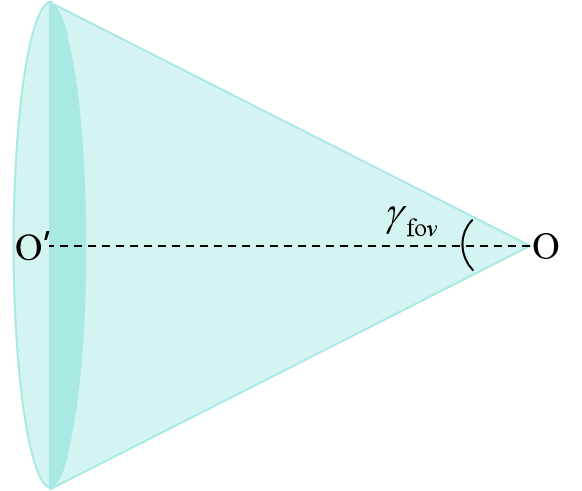}
        \caption{field of view}
    \end{subfigure}
    \caption{Illustration of the pixels on the screen and field of view of the camera}
    \label{fig:2}
\end{figure}

For simplicity, we set the energy of the observed photon to unity, i.e., $
E_{\rm camera} = 1$.
Therefore, as shown in Fig.\ref{fig:1}, in the observer’s frame, the tangent vector to the photon’s geodesic can be expressed as follows,
\bea
\dot{s} = -{e}_{0} + \cos\alpha{e}_{1}+ \sin\alpha\cos\beta{e}_{2}+ \sin\alpha\sin\beta{e}_{3}.
\eea
The negative sign in the above expression indicates that the tangent vector points to the past. 
By expressing the tangent vector as a linear combination of the coordinate basis vectors, it can also be written as
\bea
\dot{s} = \dot{t} \partial_{t}+ \dot{r} \partial_{r}+ \dot{\theta} \partial_{\theta}+ \dot{\phi} \partial_{\phi}.
\eea 
Therefore, once the four-momentum of the photon is determined, its corresponding position on the screen can also be determined.
More specifically, we choose the origin $O'$ on the screen and establish a standard Cartesian coordinate system \cite{Li:2024ctu} at this point. Based on the geometric relationships, the projection coordinates of point $P$ onto this plane can then be determined as
\bea 
x_p = - 2\tan\left(\frac{\alpha }{2}\right) \sin \beta ,\quad y_p = - 2 \tan\left(\frac{\alpha }{2}\right) \cos \beta.
\eea 
To implement this projection, it is necessary to set the camera's field of view angle $\gamma$ and define the corresponding pixel elements on the screen. 
As presented in Fig.\ref{fig:2}, for a square screen with side length $D$, its relationship with the field of view angle can be expressed as follows,
\bea 
D = 2 \tan \frac{\gamma _{\text{fov}}}{2}.
\eea
If the screen contains $n$ × $n$ pixels, the side length of each pixel, $d$, is
\bea
d = \frac{2}{n} \tan \left( \frac{\gamma _{\text{fov}}}{2} \right).
\eea
In Fig.\ref{fig:2}, the pixels are labeled as $(i, j)$, with both indices ranging from $1$ to $n$. 
The pixel in the lower-left corner of the image is labeled $(1, 1)$, 
while the pixel in the upper-right corner is labeled $(n, n)$. Thus, we have
\bea
x_p = d \left( i - \frac{n + 1}{2} \right), \quad y_p = d \left( j - \frac{n + 1}{2} \right).
\eea
Finally, the mapping between the celestial-sphere coordinates $(\alpha, \beta)$ and the pixel coordinates $(i, j)$ can be expressed as
\bea
&&\tan \beta= \frac{2j-(n + 1)}{2i-(n + 1)},\\
&&\tan \frac{\alpha}{2}=\frac{1}{n} \tan \left(\frac{\gamma _{fov}}{2}\right) \sqrt{\left(i - \frac{n + 1}{2}\right)^{2}+\left(j-\frac{n + 1}{2}\right)^{2}}.\nn 
\l. \r.
\eea 
Therefore, based on the aforementioned camera model, we can inversely trace light rays from screen pixels to determine their intersection points with the light source near the black hole, thereby reconstructing the final image of the light source around the black hole as displayed on the screen. In the next section, we will incorporate a specific light source model to numerically simulate the images of the black hole.

\section {The thin disk image of the rotating Lorentz-Violating black hole}\label{sec4}
In this section, we consider a thin accretion disk located at the equatorial plane of the rotating LV black hole. It is important to note that the motion of matter within the disk is assumed to be divided into two distinct regions by the ISCO. Specifically, outside the ISCO, the accretion flows move along time-like circular orbits, and the equation of motion of the accretion flows can be written as
\bea
u^{\mu}=u_{\text{out}}^{t}(1,0,0,\Omega_{s}),
\eea
when $\theta ={\pi }/{2}$, we obtain
\bea
u_{\text{out}}^{t}=\sqrt{-\frac{1}{g_{\phi\phi}\Omega_{s}^{2}+2g_{t\phi}\Omega_{s}+g_{tt}}}, \quad \Omega_{s}=\frac{-\partial_{r}g_{t\phi}+\sqrt{(\partial_{r}g_{t\phi})^{2}-\partial_{r}g_{\phi\phi}\partial_{r}g_{tt}}}{\partial_{r}g_{\phi\phi}}.
\eea
Then, within $ISCO$, the matter falls from the $ISCO$ to the event horizon on a critical plunging orbits with the same energy and angular momentum as those at the $ISCO$. The components of four-velocity can be given by
\bea
&&u_{in}^{t} = -g^{tt}E_{ISCO} + g^{t\phi}L_{ISCO}\big|_{\theta = \frac{\pi}{2}}, \quad u_{in}^{r} = -\sqrt{\frac{-{V}_{eff}(r)}{g_{rr}}}, \\&&u_{in}^{\phi} = -g^{t\phi}E_{ISCO} + g^{\phi\phi}L_{ISCO}\big|_{\theta = \frac{\pi}{2}}, \quad u_{in}^{\theta} = 0.\nn 
\l. \r.
\eea
Here, the minus sign in front of the square root indicates the ingoing motion of the flows. Moreover, the term of ${V}_{eff}(r)$ is the effective potential, which is 
\bea
{V}_{eff}(r)=1 + g^{tt}E^{2}+g^{\phi\phi}L^{2}-2g^{t\phi}EL.
\eea
According to the following conditions
\bea
V_{eff} = 0,\quad {\partial_r V_{eff}} = 0, \quad \partial_{r}^{2} V_{{eff}} = 0,
\eea
the position of the $ISCO$ is obtained from the effective potential, with the stability of circular orbits requiring $\partial_{r}^{2} V_{{eff}} \geq 0$.
Based on \cite{Gralla:2019xty}, the detectable light intensity depends on the number of times the light ray passes through the thin disk, as well as on the emission location relative to the observer. In this work, we neglect the refractive effects of the thin disk. The variation in light intensity along the photon's path is governed by the following relation, as given in \cite{Lindquist:1966igj}
\bea
\frac{d}{d\lambda} \left( \frac{I_{\nu}}{\nu^3} \right) = \frac{j_{\nu} - \alpha_{\nu} I_{\nu}}{\nu^2},
\eea
where $j_{\nu}$ represents the emissivity, $\alpha_{\nu}$ is the absorption 
coeffient at the frequency $\nu$, and $I_{\nu}$ is the specific intensity. In a vacuum, both $\alpha_{\nu}$ and $j_{\nu}$ are zero, which implies that ${I_{\nu}}/{\nu^3}$ is a conserved quantity along the photon geodesic. In this paper, we assume the thin disk is axisymmetric, stable, and possesses $Z_{2}$ symmetry about the equatorial plane. And, given that the thin disk considered is geometrically thin, the absorption coefficient and emissivity of light generally remain unchanged as the light passes through the disk. Considering the redshift factor $g_{n}={\nu_{o}}/{\nu_{n}}$, the total light intensity observed can be simplified to the following expression,
\bea\label{e4.7}
I_{\nu_{o}} = \sum_{n}^{n_{\text{max}}} f_{n} \left( \frac{\nu_{o}}{\nu_{n}} \right)^{3} J_{n}.
\eea
Combined with the EHT images were obtained at an observational wavelength of 1.3 mm (230 GHz), we adopt the following specific form for the emissivity,
\bea
J = \exp\left[-\frac{1}{2}\left(\log\frac{r}{r_{+}}\right)^{2} - 2\left(\log\frac{r}{r_{+}}\right)\right],
\eea\label{4.8}
where ${r_{+}}$ denotes the horizon radius. 
According to \cite{Hou:2022eev}, it is obvious that 
$\Delta I_{\nu_o} = g_n^3 \Delta I_{\nu_n} =g_n^3 J_n \Delta \tau_n = g_n^3 J_n (\nu_n \Delta \lambda_n) = f_n g_n^3 J_n$. And $f_n$ is called the “fudge factor”, which is $f_n = \nu_n \Delta \lambda_n = \Delta \tau_n$.
So, the fudge factor denotes the effective proper time interval experienced by photons as they traverse the disk plane, which serves as a local weight in the radiative-transfer integral under the zero-thickness thin-disk approximation.
In the commonly used geometric units and with the standard affine-parameter normalization, one has
$\nu_n \sim 1/\mathrm{length}$ and $\Delta\lambda_n \sim \mathrm{length}$, so that $f_n$ is dimensionless.
Its specific value depends on the microphysical structure of the emitting layer in the disk.
In this sense, the parameter $f_{n}$ mainly influences the brightness of the narrow photon ring with negligible effect on the overall image, and we can simply set $f_{n} = 1$ in this paper. 
In this work, the LV effects are mainly encoded in the parameter $\ell$.
Although $\ell$ itself does not affect the value of $f_n$, it can significantly modify
the photon redshift factor $g_n$, as well as the disk emission model. From the relation
$I_{\nu_o}=\sum_{n=1}^{n_{\text{max}}} f_n g_n^3 J_n$,
it follows that thin-disk images of rotating LV black holes can provide an intuitive manifestation of the observational features induced by LV effects.
When $r_{n}> r_{ISCO}$, the redshift factor takes
\bea
g_{out}=\frac{e}{\zeta\left(1+\frac{p_{\phi}}{p_{t}}\Omega_{n}\right)}, \quad r_{n} \geq r_{\text{ISCO}},
\eea
when $r_{n}< r_{ISCO}$,
\bea
g_{in} = -\frac{e}{u_{in}^{r}\frac{p_{r}}{p_{t}} + E_{ISCO}(g^{tt} - g^{t\phi}\xi) + L_{ISCO}(g^{\phi\phi}\xi - g^{t\phi})}, \quad r_{n} < r_{ISCO},
\eea
with
$\Omega_{n}=\left(\frac{u^\phi}{u^t}\right)|_{r = r_{n}}, \xi = \sqrt{\frac{-g_{\phi\phi}}{g_{tt} g_{\phi\phi} - g_{t\phi}^2}}, \zeta = \sqrt{\frac{-1}{g_{tt} + 2g_{t\phi}\left(\frac{u^\phi}{u^t}\right) + g_{\phi\phi}\left(\frac{u^\phi}{u^t}\right)^{2}}}$. And, $e ={p_{(t)}}/{p_{t}}$ denotes the ratio between the energy observed on the screen and the energy propagated along the null geodesic, where $p_{(\mu)} = p_{\nu} e^{\nu}_{(\mu)}$.
For the  asymptotically flat spacetimes, since the observer is located at infinity, we can set $e = 1$.

\begin{figure}[htbp]
  \centering
  \begin{subfigure}[t]{0.23\textwidth}
    \includegraphics[width=\textwidth]{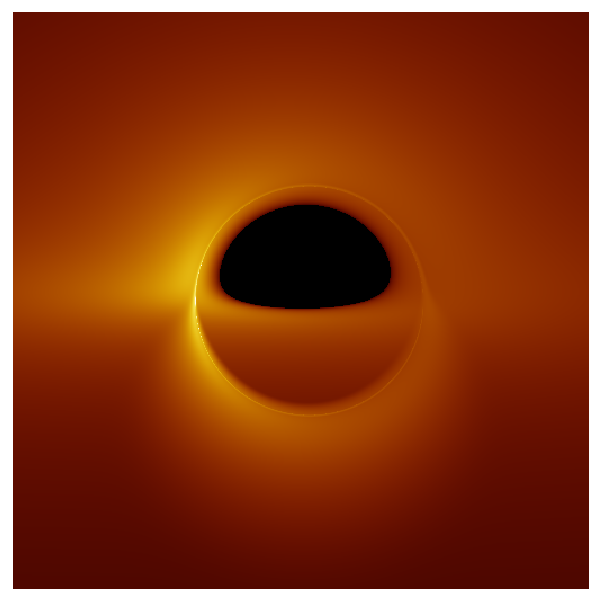}
    \caption{$a=0.1, \ell=0.99$}
  \end{subfigure}\begin{subfigure}[t]{0.23\textwidth}
    \includegraphics[width=\textwidth]{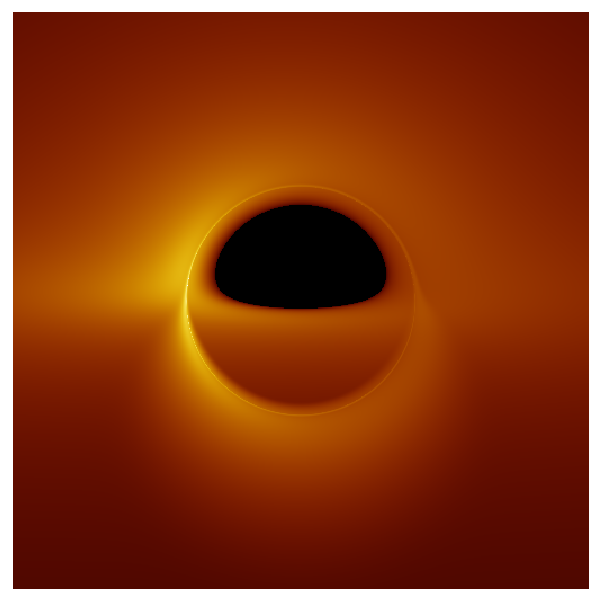}
    \caption{$a=0.1, \ell=-0.99$}
     \label{fig:3b}
  \end{subfigure}\begin{subfigure}[t]{0.23\textwidth}
    \includegraphics[width=\textwidth]{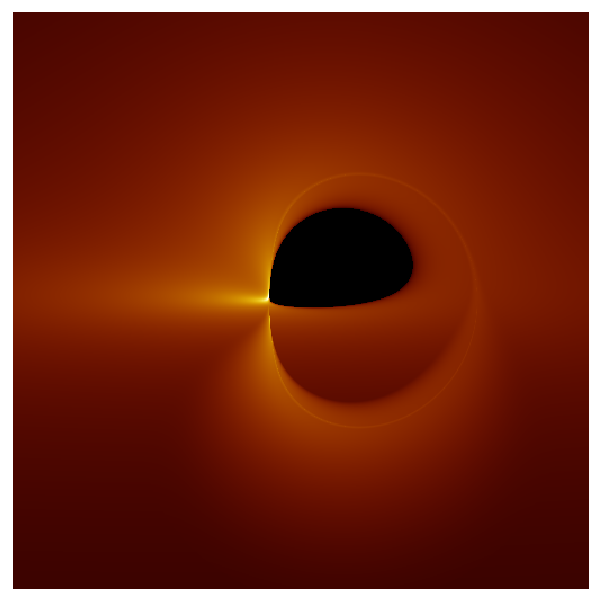}
    \caption{$a=0.94, \ell=0.99$}
  \end{subfigure}\begin{subfigure}[t]{0.23\textwidth}
    \includegraphics[width=\textwidth]{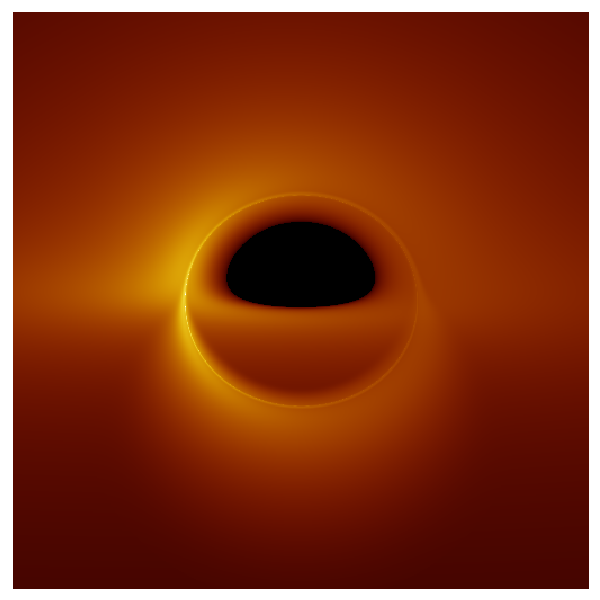}
    \caption{$a=0.94,\ell=-0.99$}
     \label{fig:3d}
  \end{subfigure}
  \caption{Images of the rotating LV black hole illuminated by the thin disk in Ho\v{r}ava gravity.}
  \label{fig:3}
\end{figure}

Based on the thin-disk model described above, we perform ray-tracing simulations to obtain the images of the rotating LV black hole. Specifically, for the observer’s position, we consistently choose $r_o=500M$ to present the images of the rotating black hole in Ho\v{r}ava gravity. Since the emission profile of the thin-disk drops off rapidly with increasing radius, the contribution from regions far from the event horizon to the observed image of the black hole is negligible. Consequently, we set the outer radius of the thin-disk to $r_{{or}} = 60M$, while the inner edge is located at the event horizon, i.e., $r_{{ir}} = r_+$. 
At present, based on the {EHT} observations, people have discussed the parameters of rotating
LV black holes and derived corresponding constraints.
In general, the parameter $\ell$ can be considered within the interval $\ell\in(-1,1)$.
Recently, one has focused mainly on one branch, namely $\ell\in[0,1)$, which corresponds to $1<\sqrt{\kappa\xi} \le 2$ \cite{Liu:2025lwj}.
Within this branch, for a given observed angle $\theta_o$, one scans the parameter grid $\ell\in[0,1)$ and $a/M\in[0,1)$, computes the shadow angular diameter point by point, and identifies those points falling within the {EHT}-measured diameter and its uncertainty band as the allowed region, thereby enabling a direct comparison with observations and parameter constraints.
For M87$^\ast$, the authors adopt the shadow angular diameter $42\pm 3~\mu{\rm as}$ and take $\theta_o=17\pi/180$. They find that in the low-spin regime the {EHT} data provide almost no effective constraint on $\ell$; however, the upper bound on the spin becomes weaker once $\ell$ is allowed: for $\ell=0$ one has approximately $a/M\lesssim 0.56$, while allowing $\ell\neq 0$ relaxes this to $a/M\lesssim 0.7$\cite{Liu:2025lwj}.
For Sgr~A$^\ast$, the authors adopt the angular diameter $51.8\pm 2.3~\mu{\rm as}$ and consider $\theta_o=\pi/2$, obtaining spin-dependent upper and lower bounds on $\ell$: at low spins they derive a lower bound on $\ell$ (e.g., $\ell>0.47$ for $a/M=0.1$ and $\ell>0.02$ for $a/M=0.5$), whereas at high spins they obtain an upper bound (e.g., $\ell<0.72$ for $a/M=0.99$). Moreover, in the range $a/M\simeq (0.6\text{, }0.8)$, current observations impose essentially no meaningful constraint on $\ell$\cite{Liu:2025lwj}.
When $\theta_{o} = 80^\circ$, the numerical results are shown in Fig.\ref{fig:3}, with the LV parameter $\ell$ set to $0.99$ and $-0.99$, respectively\footnote{In order to more clearly observe the impact of LV effect on black hole images, this study adopts relatively large parameter values.}
\begin{minipage}[t]{0.23\textwidth}
\end{minipage}

When the parameter $\ell=0.99$, a comparison between the first and third figures shows that, for a small spin parameter $a = 0.1$, the inner shadow of black hole is approximately a black semicircle, and the corresponding critical curve is nearly circular. 
In contrast, the inner shadow exhibits a slightly left-shifted and flattened structure, with a sharper and more distinct left boundary, while the right boundary appears smoother and more rounded. Meanwhile, the critical curve acquires a characteristic ``D"-shaped distortion. 
Notably, as $\ell=-0.99$, it shows in subfigures \ref{fig:3b} and \ref{fig:3d} that when $a$ increases, the shape of the inner shadow remains nearly unchanged, exhibiting only a slight overall shrinkage, while the critical curve gradually evolves from an approximately circular to an elliptical form.
When $a = 0.1$, the shapes of the inner shadow and the critical curve exhibit almost no noticeable change as the LV parameter decreases from $0.99$ to $-0.99$. 
However, when $a = 0.94$, both the inner shadow and the critical curve undergo significant changes, and their overall morphology appears to revert to the pattern observed for $a = 0.1$.
This indicates that the rotating effect of the black hole is gradually canceled as the parameter $\ell$ decreases, resulting in a critical curve that becomes more circular and an inner shadow that gradually approaches a semicircular shape. 
In next, we further present the redshift distribution maps of the direct and lensing images corresponding to Fig.\ref{fig:3}.
\begin{minipage}[t]{0.23\textwidth}
\end{minipage}
\begin{figure}[htbp]
  \centering
  \begin{subfigure}[t]{0.23\textwidth}
    \includegraphics[width=\textwidth]{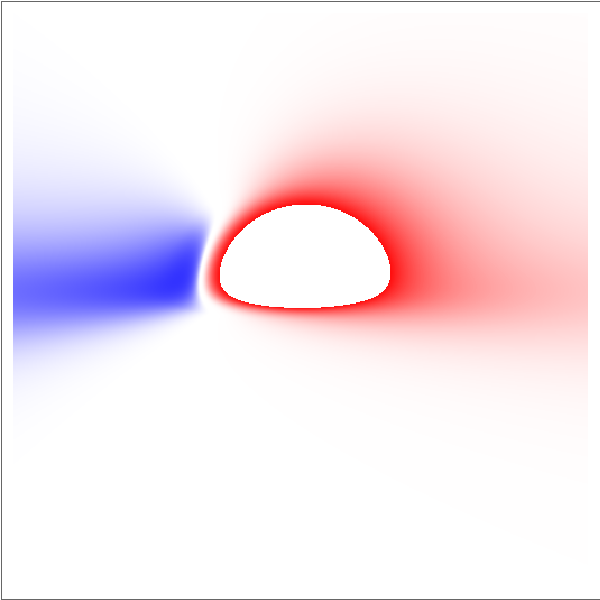}
    \caption{$a=0.1, \ell=0.99$}
    \label{fig:4a}
  \end{subfigure}\begin{subfigure}[t]{0.23\textwidth}
    \includegraphics[width=\textwidth]{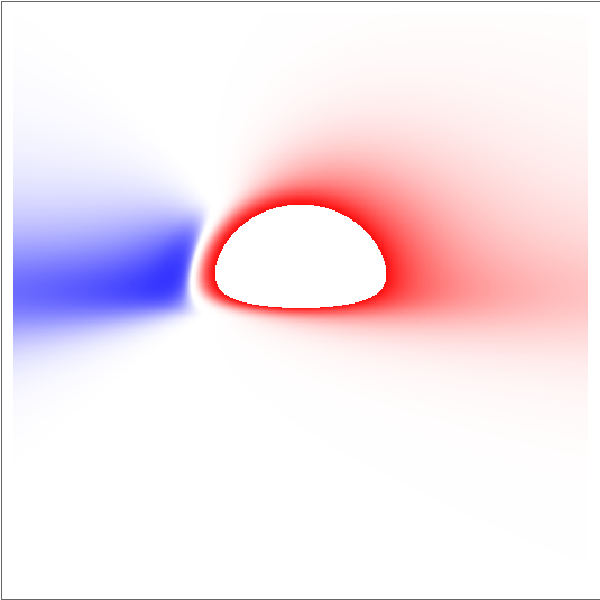}
    \caption{$a=0.1, \ell=-0.99$}
    \label{fig:4b}
  \end{subfigure}\begin{subfigure}[t]{0.23\textwidth}
    \includegraphics[width=\textwidth]{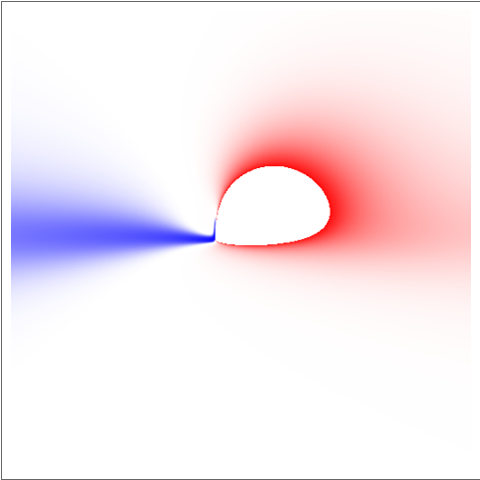}
    \caption{$a=0.94, \ell=0.99$}
    \label{fig:4c}
  \end{subfigure}\begin{subfigure}[t]{0.23\textwidth}
    \includegraphics[width=\textwidth]{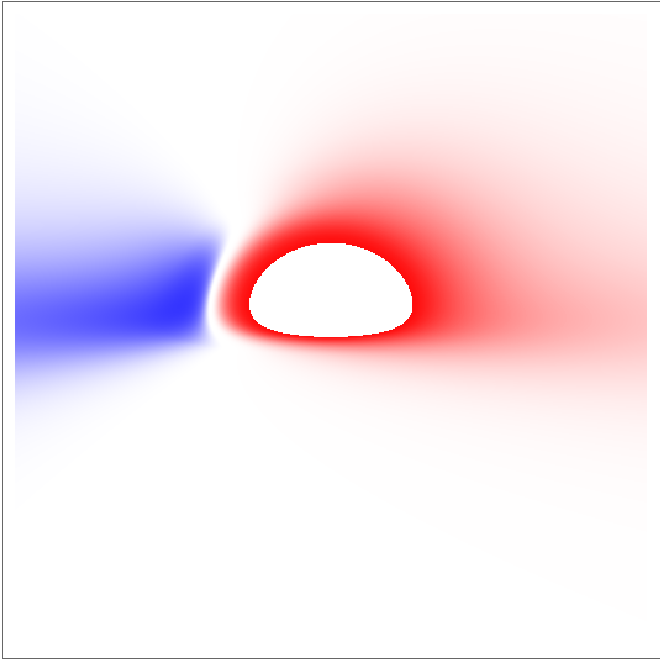}
    \caption{$a=0.94, \ell=-0.99$}
    \label{fig:4d}
  \end{subfigure}
  \caption{The redshift of direct images of the thin disk, where the red and blue color represent red shift and blue shift respectively.}
\label{fig:4}
\end{figure}
\begin{minipage}[t]{0.23\textwidth}
\end{minipage}
\begin{figure}[htbp]
  \centering
  \begin{subfigure}[t]{0.23\textwidth}
    \includegraphics[width=\textwidth]{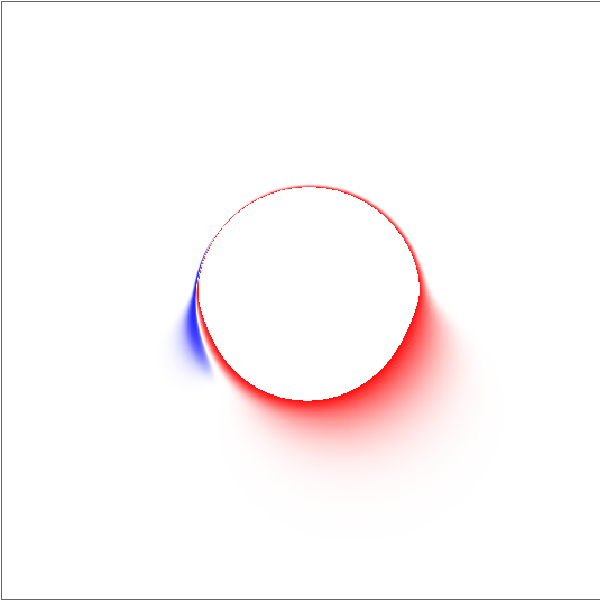}
    \caption{$a=0.1, \ell=0.99$}
  \end{subfigure}\begin{subfigure}[t]{0.23\textwidth}
    \includegraphics[width=\textwidth]{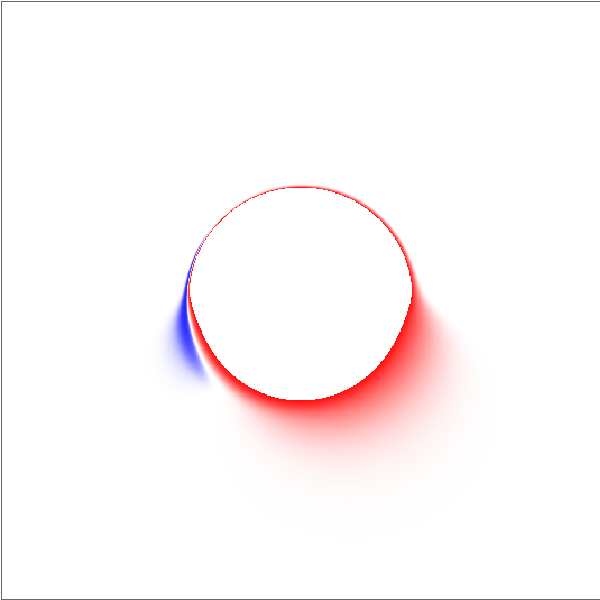}
    \caption{$a=0.1, \ell=-0.99$}
  \end{subfigure}\begin{subfigure}[t]{0.23\textwidth}
    \includegraphics[width=\textwidth]{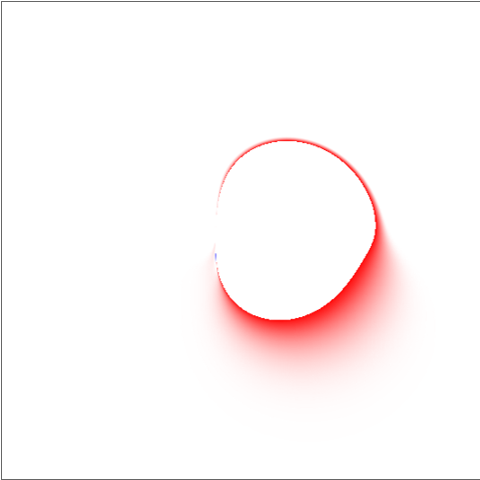}
    \caption{$a=0.94, \ell=0.99$}
  \end{subfigure}\begin{subfigure}[t]{0.23\textwidth}
    \includegraphics[width=\textwidth]{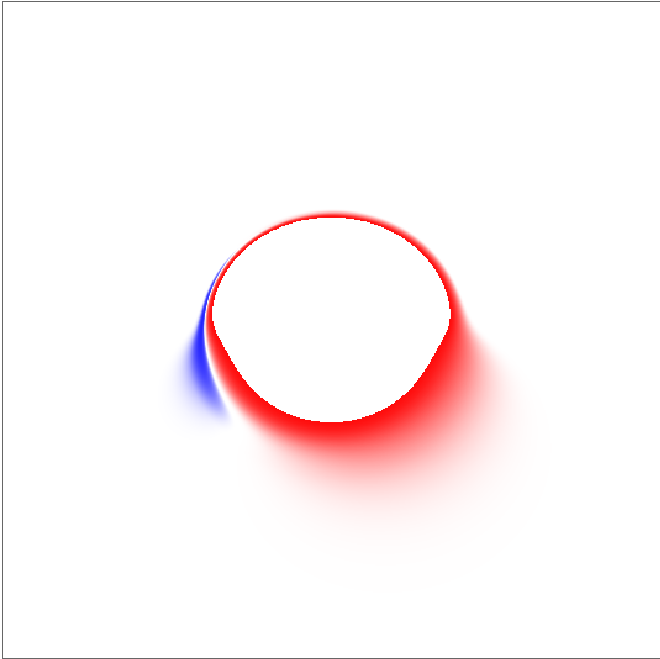}
    \caption{$a=0.94, \ell=-0.99$}
  \end{subfigure}
  \caption{The redshift of lensed images of the thin disk.}
\label{fig:5}
\end{figure}

From Fig.\ref{fig:4}, it can be seen that when the rotating parameter takes the value $a = 0.1$, variations in the LV parameter have only a mild effect on the redshifted image. In contrast, when $a = 0.94$, the situation changes significantly. As shown in subfigures \ref{fig:4c} and \ref{fig:4d}, for $\ell = 0.99$, the left edge of the inner shadow is relatively sharp, and the redshifted and blueshifted regions are narrowly distributed; whereas for $\ell = -0.99$, the left side of the inner shadow becomes more rounded and even bulges to the right, with a significantly enlarged redshifted and blueshifted area, and its overall distribution resembles that observed for $a = 0.1$.
For the rotating parameter $a$, we observe that when $\ell = 0.99$, the redshifted and blueshifted regions are more sensitive to changes in $a$, compared to the case of $\ell = -0.99$.
In Fig.\ref{fig:5}, the redshift distribution of the lensed image exhibits a trend with parameter variations similar to that of the direct image Fig.\ref{fig:4}. 
To further analyze the image features, we numerically extracted the intensity distributions along the x and y axes of the image in Fig.\ref{fig:3}, as shown below.
\begin{figure}[htbp]
  \centering
  \begin{tikzpicture}

    \node at (0,0) {
      \includegraphics[width=0.425\textwidth]{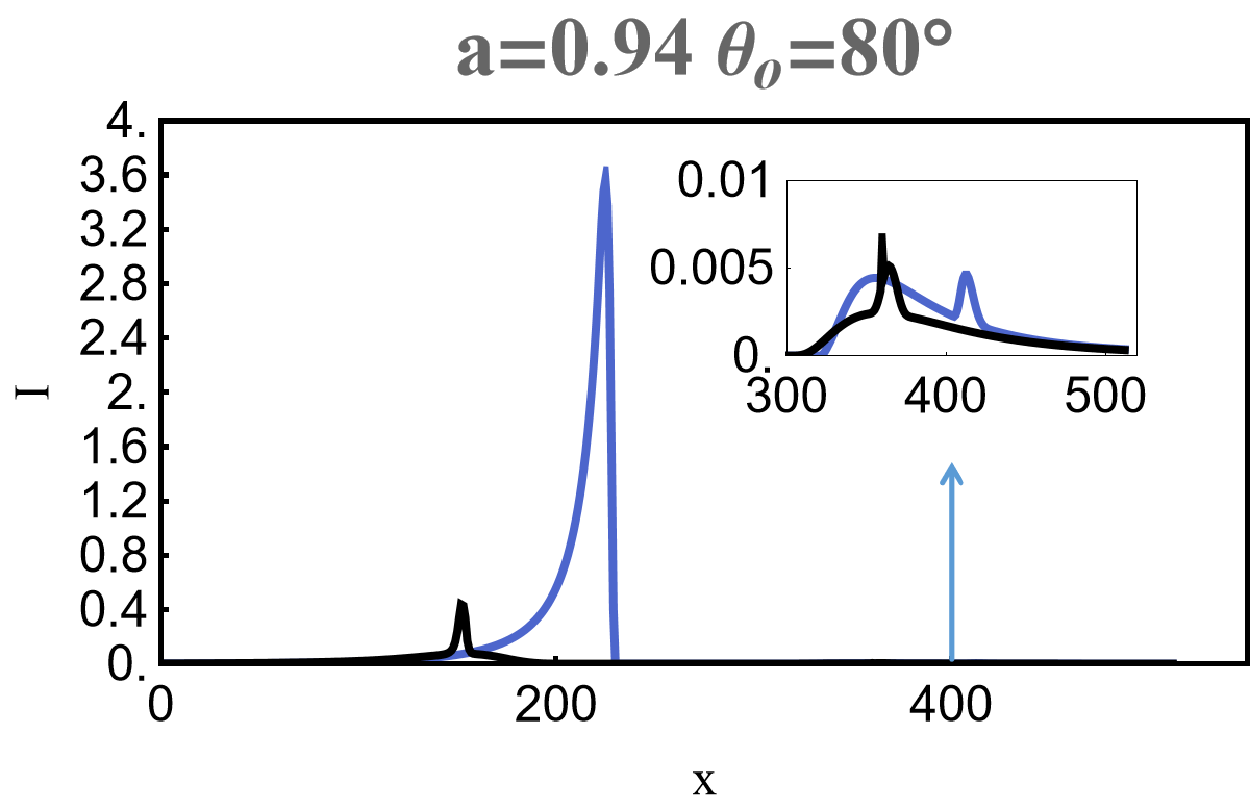}
    };
    \node at (7,-0.125) {
      \includegraphics[width=0.465\textwidth]{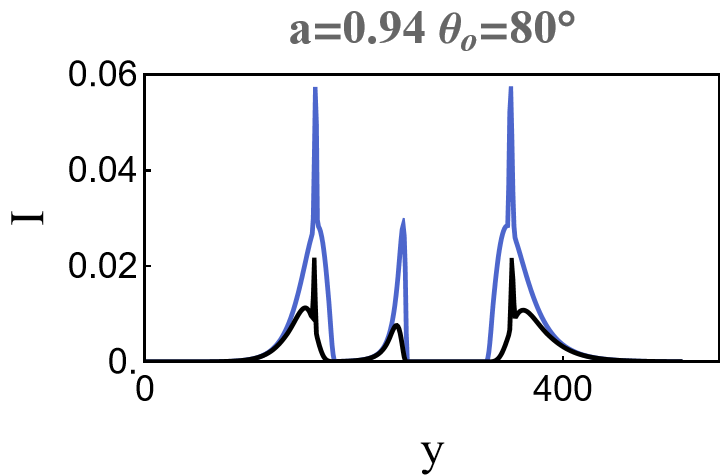}
    };
  \end{tikzpicture}
  \caption{The intensity distribution of the screen along the $x$-axis and the $y$-axis. For the first two graphs, $a=0.1$(blue line) and $a=0.94$(black line), while for the last two graphs, $\ell=0.99$(blue line) and $\ell=-0.99$(black line).}
  \label{fig:6}
\end{figure}

After analyzing the inner shadow and the critical curve, we proceed to investigate how the LV parameter $\ell$ affects the brightness distribution of the bright ring, shown in Fig.\ref{fig:6}. 
From the above images, it is evident that as the increase of LV parameter $\ell$, the peak position along the $y$-axis remains almost unchanged, but the peak intensity increases markedly. Along the $x$-axis, although the peak intensity also increases notably, the peak position shifts evidently to the right with increasing $\ell$.
Combining the above results, this indicates that the thin-disk image of the rotating LV black hole exhibits distinct and identifiable features, including the inner shadow, the critical curve, and the brightness distribution of the bright ring. These observable effects, which reflect the spacetime properties of the black hole, may offer potential means to identify and probe the LV effects.
Taking into account the polarization properties of the emitted radiation, in the next section we further analyze the influence of the LV parameter on the polarized images of the thin disk as seen by a distant observer. By jointly examining the thin disk and polarized images, we aim to gain a more comprehensive understanding of the spacetime characteristics of the rotating LV black hole.

\section{The polarized images of the rotating Lorentz-Violating black hole} \label{sec5}
\begin{figure}[htbp]
  \centering
  \begin{subfigure}[t]{0.23\textwidth}
    \includegraphics[width=\textwidth]{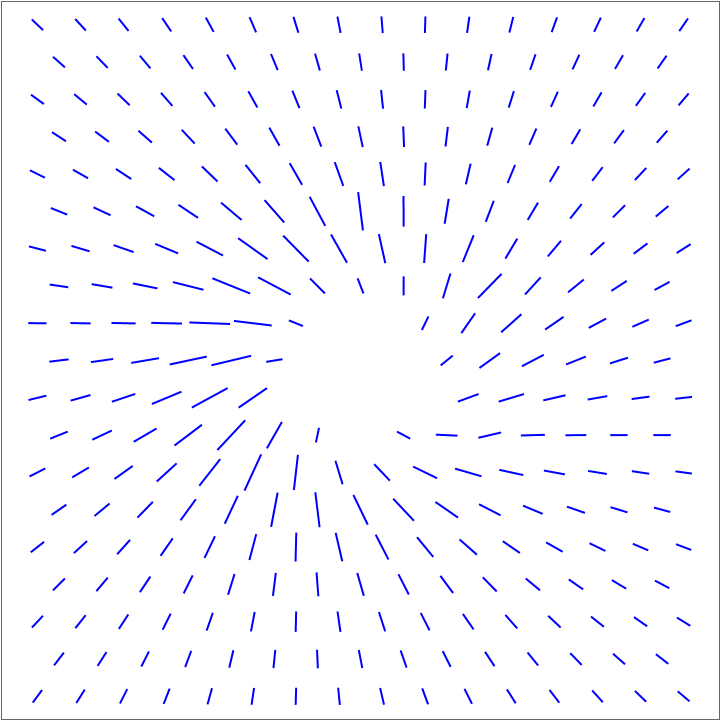}
    \caption{$a=0.1, \ell=0.99$}
    \label{fig:7a}
  \end{subfigure}
  \begin{subfigure}[t]{0.23\textwidth}
    \includegraphics[width=\textwidth]{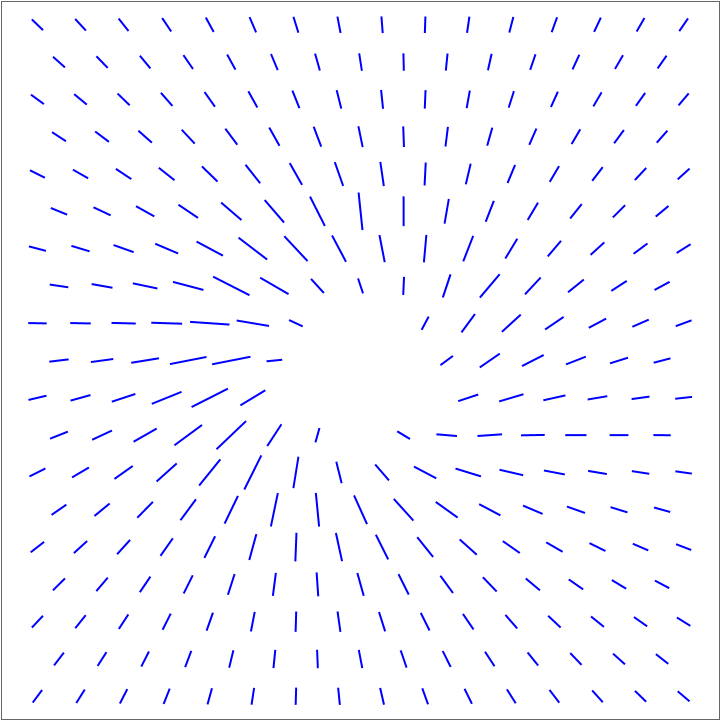}
    \caption{$a=0.1, \ell=-0.99$}
    \label{fig:7b}
  \end{subfigure}
  \begin{subfigure}[t]{0.23\textwidth}
    \includegraphics[width=\textwidth]{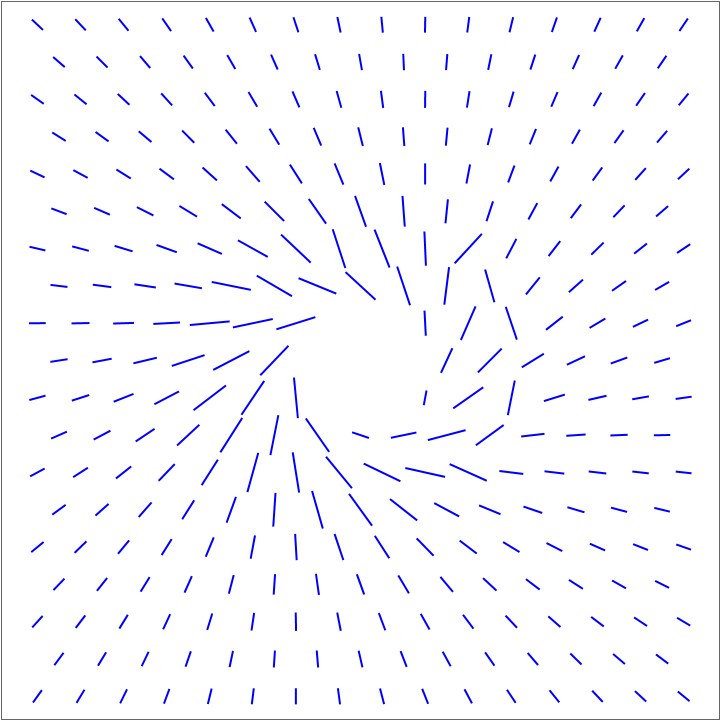}
    \caption{$a=0.94, \ell=0.99$}
     \label{fig:7c}
  \end{subfigure}
  \begin{subfigure}[t]{0.23\textwidth}
    \includegraphics[width=\textwidth]{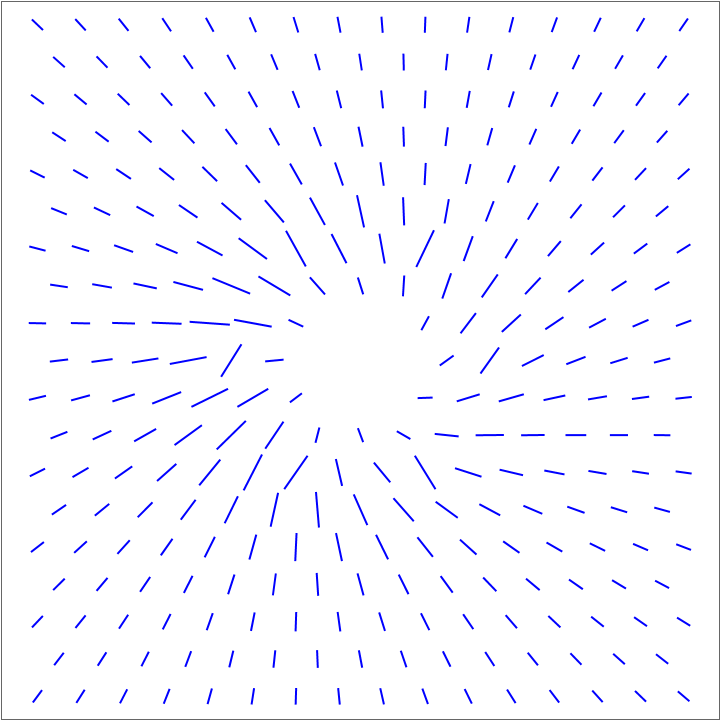}
    \caption{$a=0.94, \ell=-0.99$}
     \label{fig:7d}
  \end{subfigure}
  \caption{The polarized intensity tick plots of the rotating LV black hole with the observed position $\theta_{o} = 163^{\circ}$.}
  \label{fig:7}
\end{figure}

\begin{figure}[htbp]
  \centering
  \begin{subfigure}[t]{0.23\textwidth}
    \includegraphics[width=\textwidth]{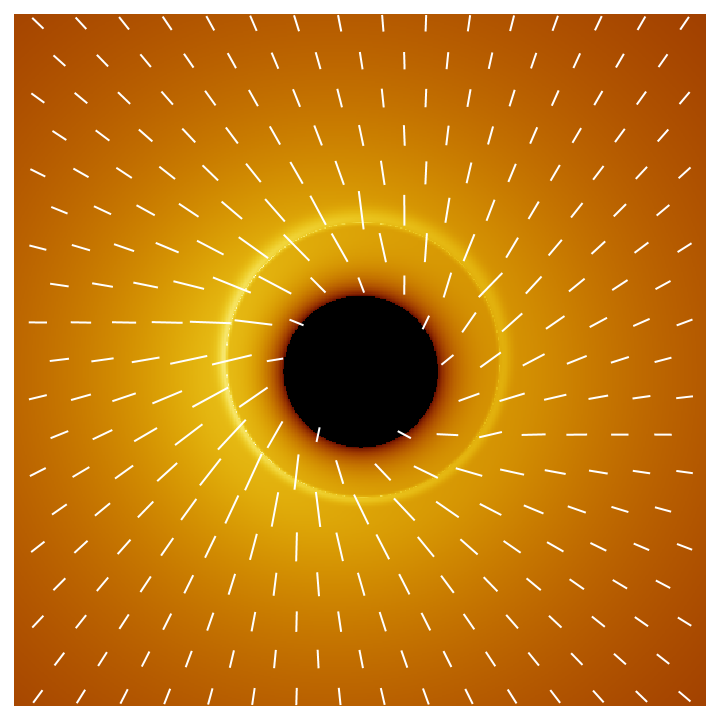}
    \caption{$a=0.1, \ell=0.99$}
      \label{fig:8a}
  \end{subfigure}
  \begin{subfigure}[t]{0.23\textwidth}
    \includegraphics[width=\textwidth]{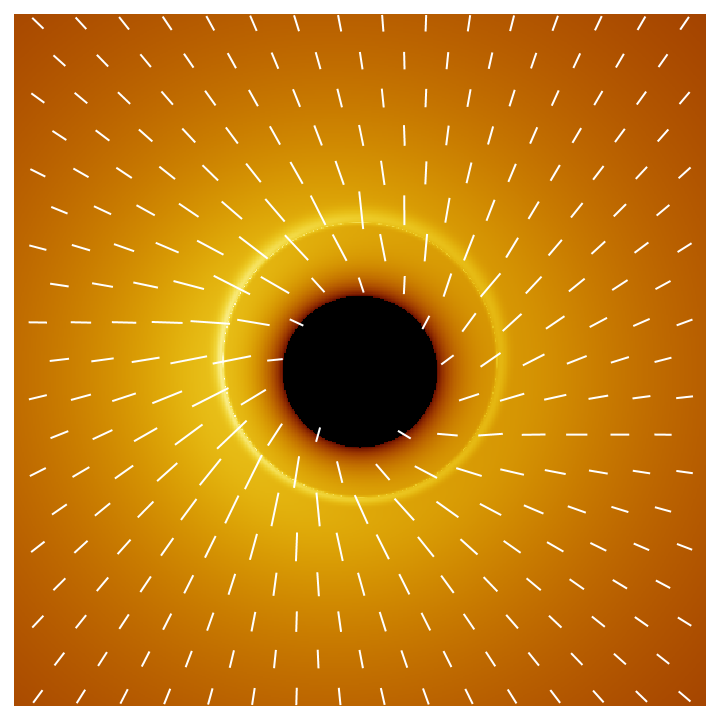}
    \caption{$a=0.1, \ell=-0.99$}
      \label{fig:8b}
  \end{subfigure}
  \begin{subfigure}[t]{0.23\textwidth}
    \includegraphics[width=\textwidth]{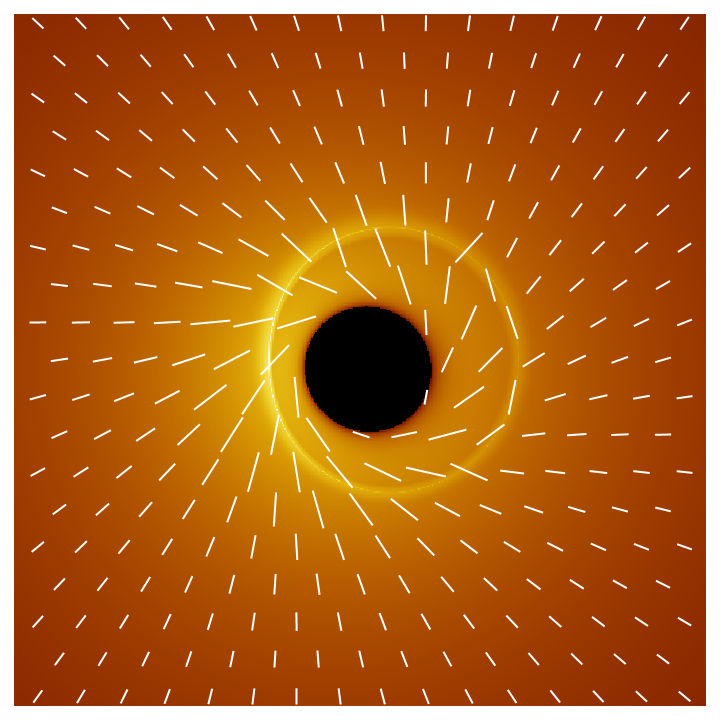}
    \caption{$a=0.94, \ell=0.99$}
      \label{fig:8c}
  \end{subfigure}
  \begin{subfigure}[t]{0.23\textwidth}
    \includegraphics[width=\textwidth]{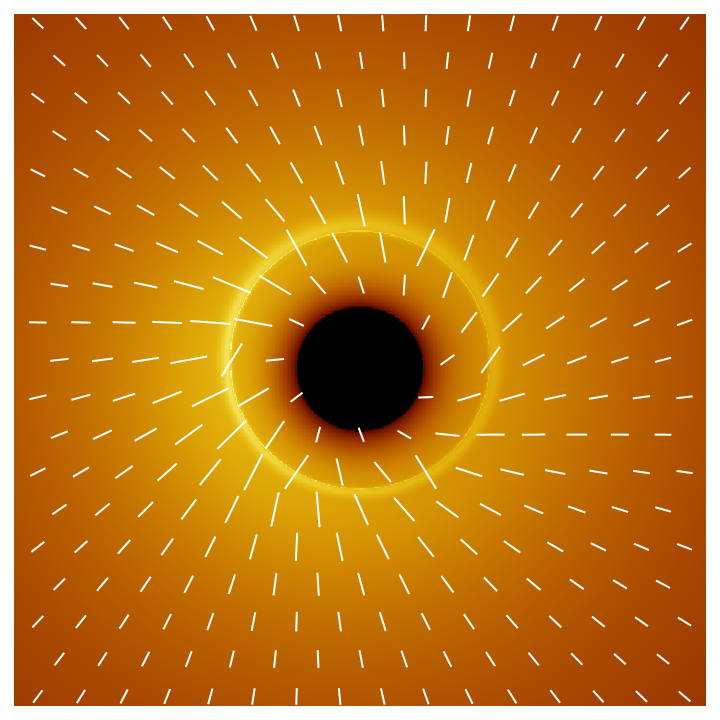}
    \caption{$a=0.94, \ell=-0.99$}
      \label{fig:8d}
  \end{subfigure}
  \caption{The polarized intensity tick plots as well as the disk image for the prograde case, where $\theta_{o} = 163^{\circ}$. }
  \label{fig:8}
\end{figure}

In this context, we assume that the emission of polarized light originates from synchrotron radiation produced by electrons within a plasma, and the observer is comoving with the plasma\cite{Li:2025awg}. Therefore, the polarization direction of the emitted light can be expressed as
\bea
\vec{f} = \frac{\vec{k} \times \vec{B}}{|\vec{k}| \, |\vec{B}|},
\eea
where $\vec{f}$ is perpendicular to both the wave vector of photon $\vec{k}$, and the local magnetic field $\vec{B}$. We now proceed to express the above relation in a generally covariant form
\bea
f^{\mu} \propto \epsilon^{\mu \nu \rho \sigma} u_{\nu} k_{\rho} B_{\sigma}.
\eea
Here, $u^\mu$ is the four-velocity of the fluid, $k^\mu$ represents the four-wavevector of the photon, and $B^\mu$ is the magnetic field.
In general, the direction of the polarization vector can be determined first, followed by normalization to ensure it satisfies the orthonormal condition, that is, $f^\mu f_\mu = 1$. The intensities of linearly polarized and natural light at the emission site are represented by the emission functions $J_p$ and $J_i$, respectively. To simplify the model, we assume that the emission intensity depends only on position, with no dependence on photon frequency or the local magnetic field, that is
\bea
J_i = J_i(r), \qquad J_p = \eta J_i(r).
\eea
In the above equation, $\eta$ denotes the ratio of linearly polarized light to the total light intensity at the point of emission, and $\eta$ lies within the interval [0, 1]. Assuming the emission is entirely linearly polarized, the value of parameter $\eta$ is taken as $\eta=1$. According to the geometric optics approximation, the polarization vector $f^\mu$ is subject to parallel transport along the photon's geodesic trajectory
\bea
k^{\nu} \nabla_{\!\nu} f^{\mu} = 0.
\eea
If the affine parameter $\lambda$ is taken into account, the above expression can be written as
\bea
\frac{\mathrm{d}}{\mathrm{d}\lambda} f^{\mu} + \Gamma_{\!\nu\rho}^{\mu} k^{\nu} f^{\rho} = 0.
\eea
When observed from the observer's position, similar to the unpolarized case, the intensity $P_{\nu o}$ of linearly polarized light and the total intensity $I_{\nu o}$ can be expressed by the following equation
\bea
P_{\nu o} = g^{3} J_{p}, \quad I_{\nu o} = g^{3} J_{i},
\eea
where $g$ is redshift factor. As described in Section \ref{sec3}, by using the ZAMO frames and imaging screen at the observer's location, the projection of the polarization vector onto the coordinate plane can be expressed as
\bea
f^{(\alpha)} = f^{\mu} \cdot e_{\alpha} = -f^{\mu} \cdot e_{\varphi}, \quad 
f^{(\beta)} = f^{\mu} \cdot e_{\beta} = -f^{\mu} \cdot e_{\theta}.
\eea
It should be noted that the two conditions $f(\beta) > 0$ and $\chi \in (0, \pi)$ need to be followed. According to the definitions of the Stokes parameters $Q$ and $U$, these parameters conform to the principle of linear superposition. Once the direction of the polarization vector is determined, the total intensity of the linearly polarized light observed by the observer can be obtained by summing the contributions from each emission point of the equatorial plane. The final result can then be obtained by directly summing the values of $Q$ and $U$, as follows
\bea
Q_{\text{obs}} = \sum_{n=1}^{n_{\text{max}}} g_{n}^{3} J_{p_{n}} \left( \left( f_{n}^{(\alpha)} \right)^2 - \left( f_{n}^{(\beta)} \right)^2 \right), \quad 
U_{\text{obs}} = \sum_{n=1}^{n_{\text{max}}} g_{n}^{3} J_{p_{n}} \left( 2 f_{n}^{(\alpha)} f_{n}^{(\beta)} \right).
\eea
Here, $n = 1, 2, 3, \ldots, n_{\text{max}}$ represents the number of times the light passes through the thin disk. Consequently, the complete intensity of the linearly polarized light is expressed as
\bea
P_{o} = \sqrt{Q_{\text{obs}}^{2} + U_{\text{obs}}^{2}}.
\eea
The electric vector position angle (EVPA) is given by
\bea
\chi = \frac{1}{2} \arctan \frac{U_{\text{obs}}}{Q_{\text{obs}}}.
\eea 
Based on the observations reported by the EHT for M87*, the magnetic field configuration $B(r, \theta, \varphi) = (0.87, 0, 0.5)$ has been shown to most accurately reproduce the observed polarization features. 
Based on the specific form of the thin-disk emissivity(\ref{4.8}), we can numerically obtain the observed polarization pattern of the linearly polarized emission on the observer’s screen, which are shown in Figs.\ref{fig:7} and \ref{fig:8}, where the Fig.\ref{fig:8} shows the result obtained by superimposing the polarization pattern onto the thin disk image.
The values of the black hole and thin-disk parameters are the same as those in Fig.\ref{fig:3}, except for the observed angle. At $\theta_o = 163^\circ$, the observer is positioned near the south pole. At this position, when $a = 0.1$, the LV parameter has little effect on the inner shadow and bright ring for Fig.\ref{fig:8}. However, for $a = 0.94$ and $\ell = 0.99$, the inner shadow becomes asymmetrically skewed to the left relative to the center. In the case of $\ell = -0.99$, the inner shadow gradually restores to the symmetric shape observed for $a = 0.1$.
\begin{minipage}[t]{0.23\textwidth}
\end{minipage}
\begin{minipage}[t]{0.23\textwidth}
\end{minipage}
\begin{minipage}[t]{0.23\textwidth}
\end{minipage}
\begin{minipage}[t]{0.23\textwidth}
\end{minipage}

\begin{figure}[htbp]
  \centering
  \begin{subfigure}[t]{0.23\textwidth}
    \includegraphics[width=\textwidth]{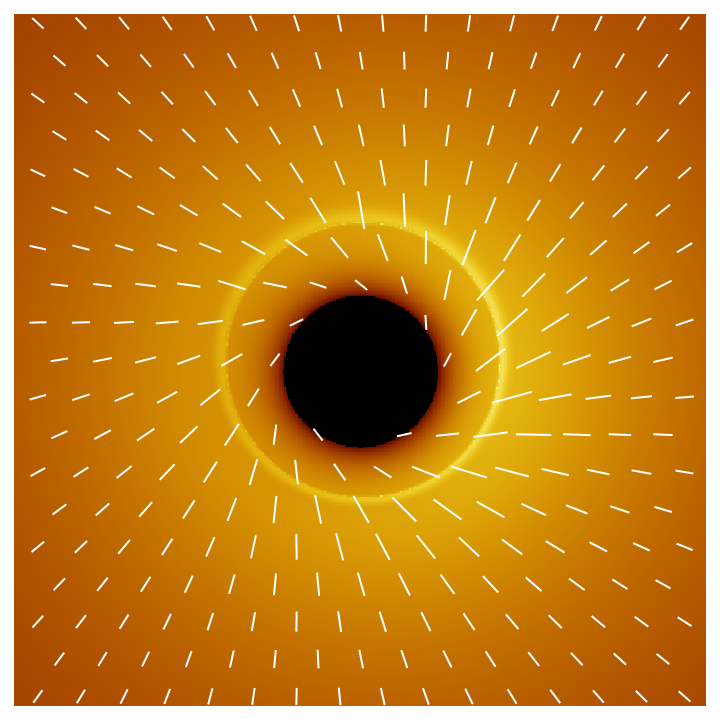}
    \caption{$a=0.1, \ell=0.99$}
  \end{subfigure}
  \begin{subfigure}[t]{0.23\textwidth}
    \includegraphics[width=\textwidth]{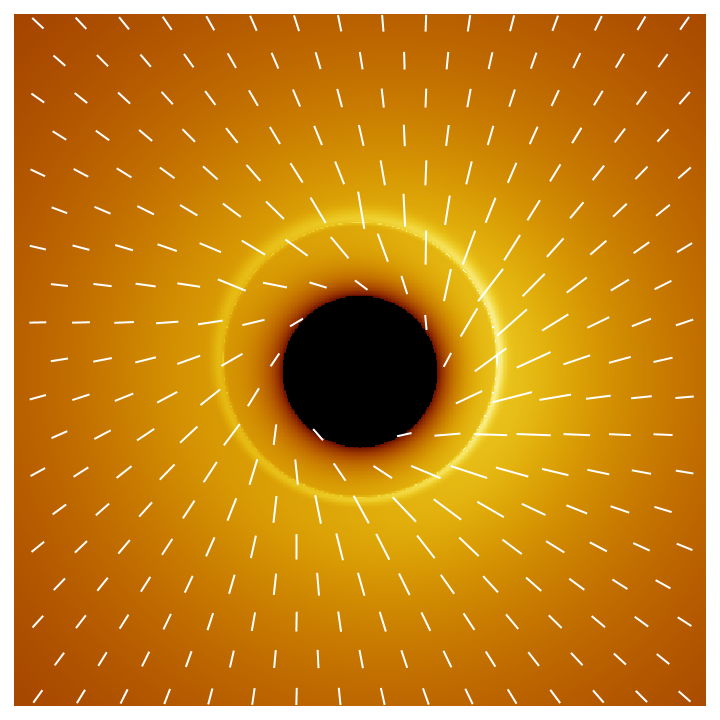}
    \caption{$a=0.1, \ell=-0.99$}
  \end{subfigure}
  \begin{subfigure}[t]{0.23\textwidth}
    \includegraphics[width=\textwidth]{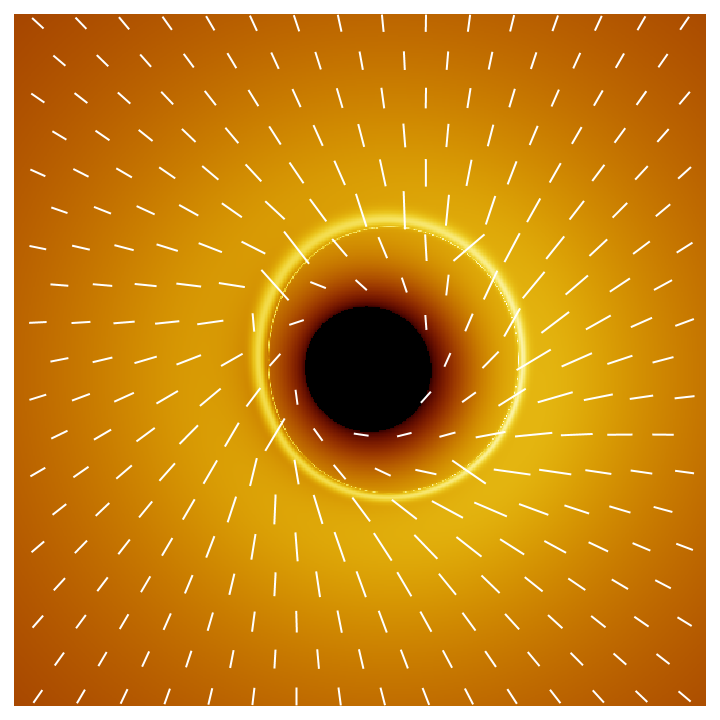}
    \caption{$a=0.94, \ell=0.99$}
  \end{subfigure}
  \begin{subfigure}[t]{0.23\textwidth}
    \includegraphics[width=\textwidth]{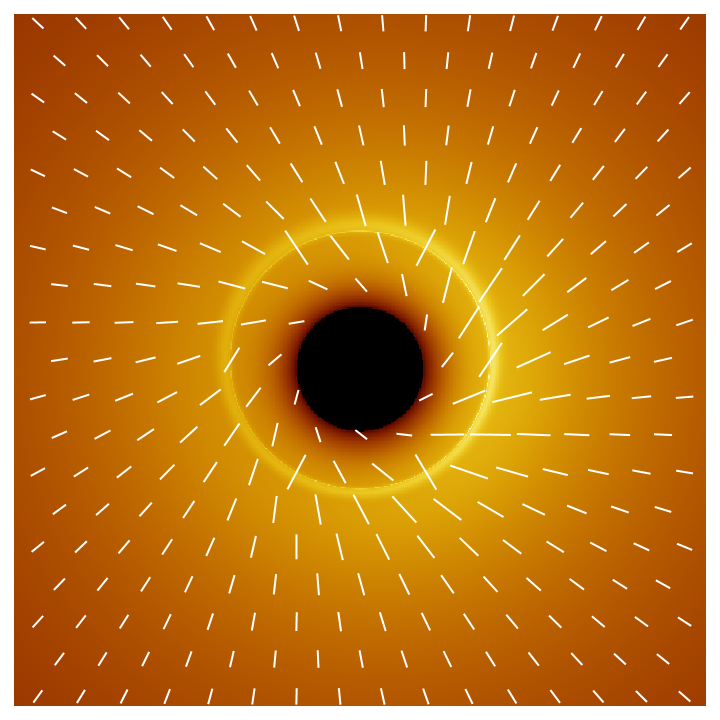}
    \caption{$a=0.94, \ell=-0.99$}
  \end{subfigure}
  \caption{The polarized intensity tick plots as well as the disk image for the retrograde case.}
  \label{fig:9}
\end{figure}

By closely examining Fig.\ref{fig:7}, we can see that a distinct central ``blank region" is visible in all four subfigures, representing the inner shadow of black hole. 
The short blue lines denote the polarization vectors, arranged in a radially divergent pattern. 
The lengths of these vectors reflect the spatial variation of the polarization intensity: longer vectors indicate stronger polarization in the corresponding region, while shorter vectors represent weaker polarization. Overall, the polarization intensity diminishes progressively with distance from the black hole.
Comparing the first two subfigures, in the case of a slowly rotating black hole, the observed polarization patterns of the two are nearly identical, indicating that the polarization pattern-including both polarization intensity and direction-depends only weakly on the LV parameter and can be almost neglected.
For a rapidly rotating black hole, it implies form the last two subfigures that the polarization pattern exhibits high sensitivity to the LV parameter, particularly in the ring-like region near the exterior of the ``blank region”, where both the polarization intensity and direction undergo significant variations, as shown in subfigure \ref{fig:7c}.
And, for subfigure \ref{fig:7d}, we further observe that when $\ell$ takes a small negative value, the black hole’s polarization pattern gradually returns to the distribution of the low-spin case.
Moreover, we also performed numerical simulations for the retrograde accretion flow, and the results are shown in the Fig.\ref{fig:9}.
\begin{minipage}[t]{0.23\textwidth}
\end{minipage}
\begin{minipage}[t]{0.23\textwidth}
\end{minipage}

Based on Fig.\ref{fig:9}, it can be seen that the thin-disk image of the black hole with a retrograde accretion flow exhibits the same inner shadow and critical curve structures as in the prograde case. The main difference between the two lies in the brightness distribution: in the prograde case, the intensity on the right side of the critical curve is higher than on the left, whereas in the retrograde case, the opposite occurs. Regarding the polarization pattern, the polarization direction undergoes significant changes near the “blank region”, while the variation in polarization intensity remains consistent with the observed brightness of the thin-disk image.
Combined with above results, it is thus evident that the polarization pattern of a black hole, as an observable effect, not only effectively reflects the spacetime structure of the black hole but also reveals the magnetic field distribution and relevant physical properties of the thin disk.

\section{The comparison between the Kerr black hole and the rotating Lorentz-Violating black hole} \label{sec6}
In Sections \ref{sec4} and \ref{sec5}, we have provided a detailed analysis of the thin-disk image and polarization characteristics of the rotating LV black hole. Based on this foundation, this section conducts a direct comparative analysis of the observable effects between the LV black hole and Kerr black hole, aiming to elucidate the marked differences in their optical images and polarization patterns, thereby quantifying the observable impact of the LV parameter $\ell$. The comparison between two black holes have been shown in following Figs.\ref{fig:10} and \ref{fig:11}. 
\begin{minipage}[t]{0.32\textwidth}
\end{minipage}
\begin{figure}[htbp]
  \centering
  \begin{subfigure}[t]{0.32\textwidth}
    \includegraphics[width=\textwidth]{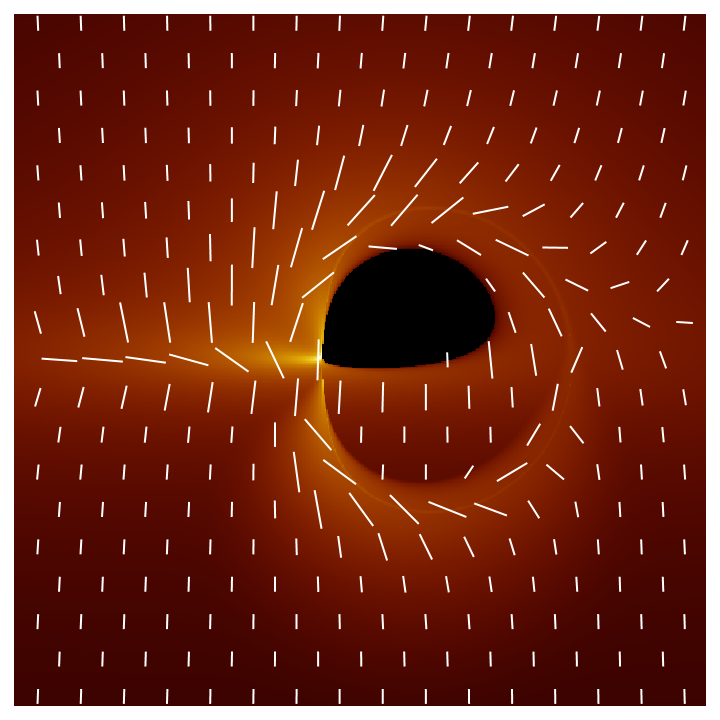}
     \caption{$a=0.94, \ell=0.99$}
     \label{fig:10a}
  \end{subfigure}
   \begin{subfigure}[t]{0.32\textwidth}
    \includegraphics[width=\textwidth]{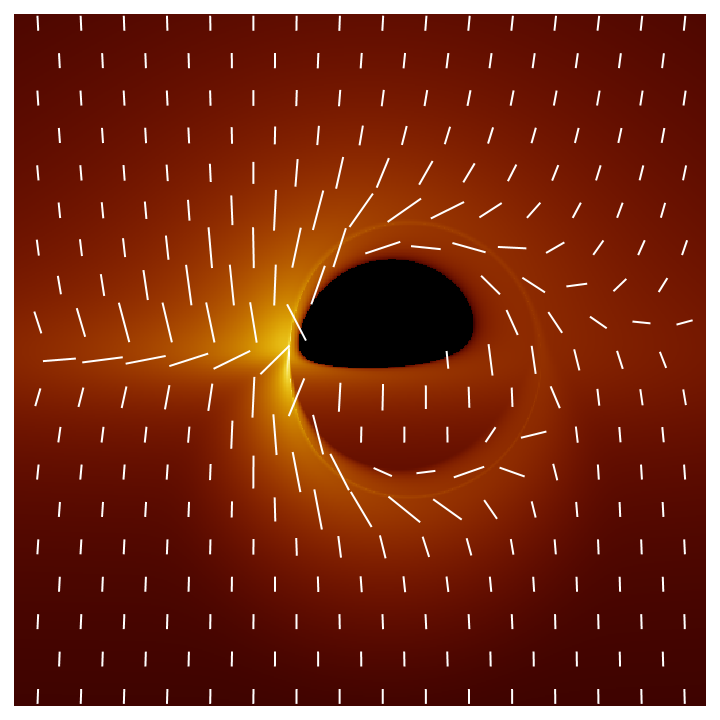}
     \caption{$a=0.94, \ell=0$}
     \label{fig:10b}
  \end{subfigure}
  \begin{subfigure}[t]{0.32\textwidth}
    \includegraphics[width=\textwidth]{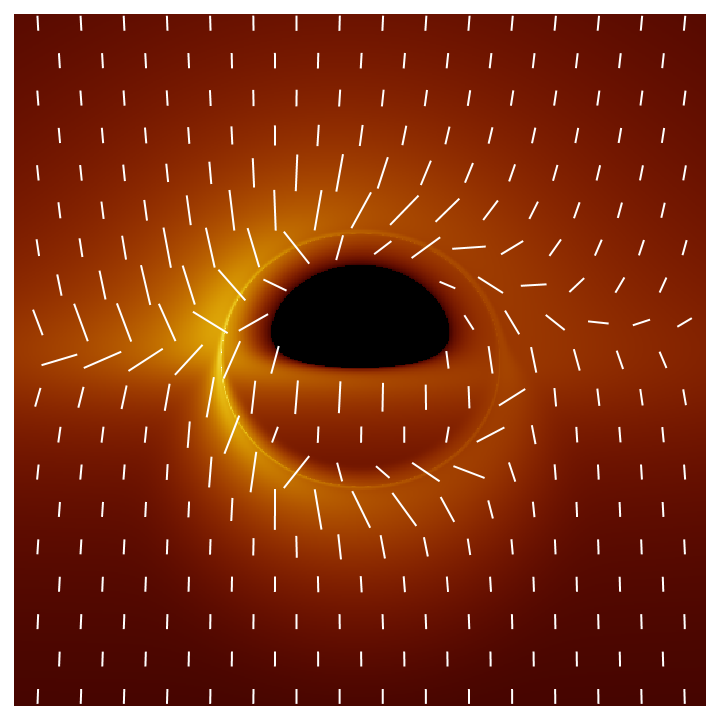}
     \caption{$a=0.94, \ell=-0.99$}
     \label{fig:10c}
  \end{subfigure}
  \caption{The observable images of Kerr black holes and LV black holes with $\theta_o=80^\circ$.}
  \label{fig:10}
\end{figure}

The case $\ell=0$ correspond to the Kerr black hole. 
At $a=0.94$, the critical curve of the Kerr black hole approximates a ``D"-shaped configuration, while its inner shadow manifests as a semi-circular structure with a slight leftward inclination. As $\ell$ decreases from 0 to $-0.99$, the critical curve of the corresponding LV black hole progressively evolves into a more elliptical form. Concurrently, the inner shadow transitions into a semi-elliptical shape and gradually loses its tilted appearance. Conversely, when $\ell$ increases to $0.99$, the critical curve adopts a configuration that more closely resembles a standard ``D" shape than in the Kerr case, accompanied by an inner shadow that exhibits a more pronounced leftward-tilted semi-circular morphology.
From Fig.\ref{fig:10}, it can be seen that the polarized  intensity is proportional to the radiative intensity of the thin disk. Consequently, the polarized intensity near the critical curve (i.e., the length of the white line segments in Fig.\ref{fig:10}) is significantly stronger than that in other regions. 
The white line segments in the figure not only represent the magnitude of the polarized intensity but also indicate the orientation of the polarization vectors. 
Near the critical curve, the polarization direction exhibits noticeable variations, whereas in regions far from black hole it remains nearly unchanged.
Compared with the Kerr case, when $\ell = 0.99$, the variation trend of the polarized intensity is generally consistent with the brightness distribution of the thin disk image. 
However, since the parameter $\ell$ affects the intensity distribution of the thin-disk image, it consequently has a direct impact on the distribution of polarized intensity.
It is worth noting that as $\ell$ increases, the distribution region of the polarized intensity changes, and the polarization direction exhibits significant differences. In other words, the polarization direction varies notably at the same spatial locations and also shows pronounced deviations near the respective critical curves. As shown in subfigures \ref{fig:10a} and \ref{fig:10b}, in the brightest regions of the image, the polarization direction of the LV black hole is clearly distinct from that of Kerr black hole. When $\ell = -0.99$, it can still be observed that the intensity of the polarization remains consistent with the brightness distribution of the thin-disk image, while the polarization direction changes as $\ell$ decreases. Compared with the case of $\ell = 0.99$, not only does the distribution of polarization intensity differ, but the polarization direction near the critical curves of the respective accretion images also exhibits clear distinctions.
To compare with the intensity distribution of the thin disk in the Kerr case, we further extracted the intensity profiles of the thin disk along the $x$- and $y$-axes and analyzed the effect of the parameter $\ell$ on the intensity distribution, which are shown in Fig.\ref{fig:11}.

\begin{figure}[htbp]
  \centering
  \begin{tikzpicture}

    \node at (0,0) {
      \includegraphics[width=0.425\textwidth]{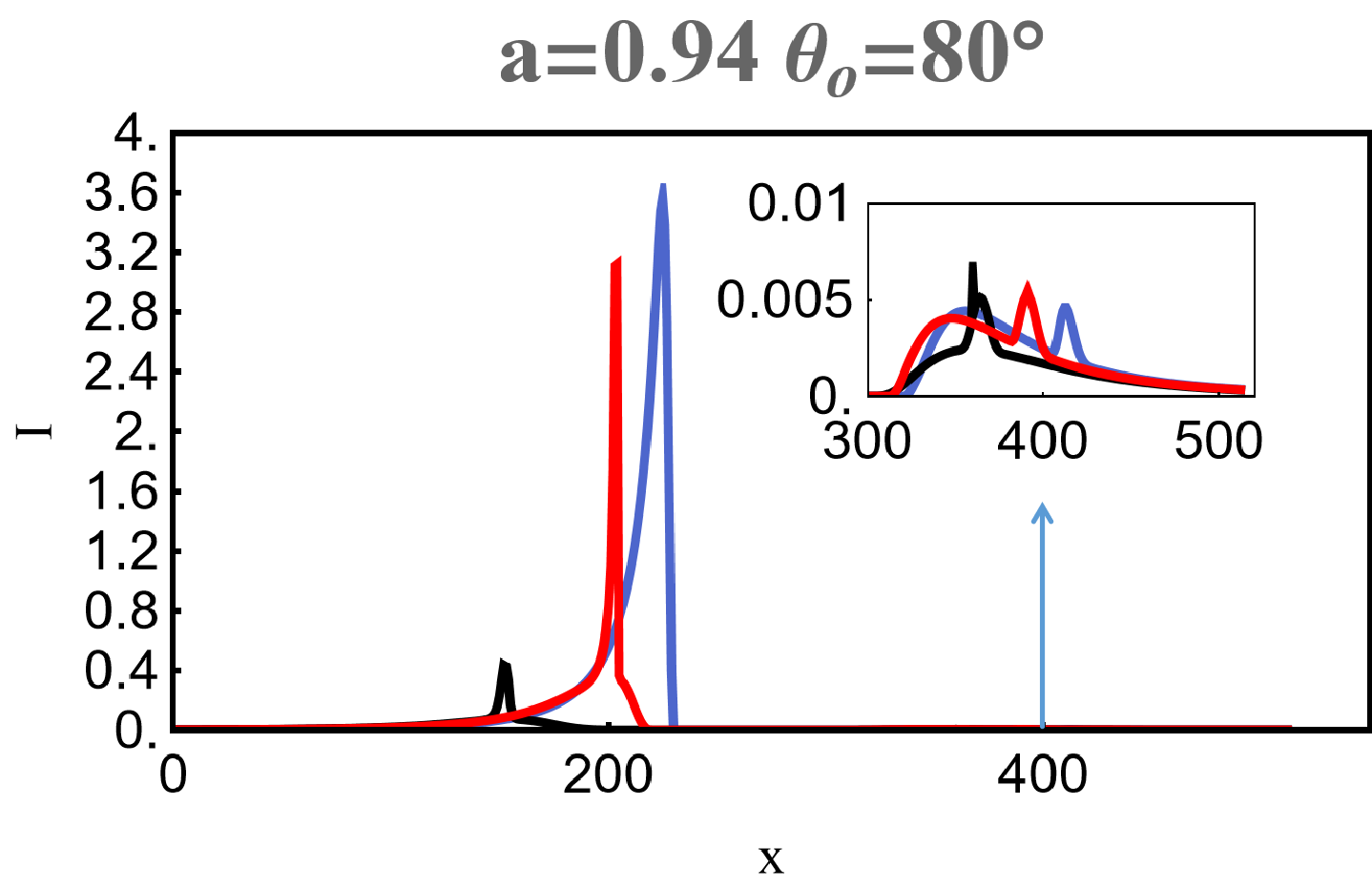}
    };
    \node at (7,-0.125) {
      \includegraphics[width=0.465\textwidth]{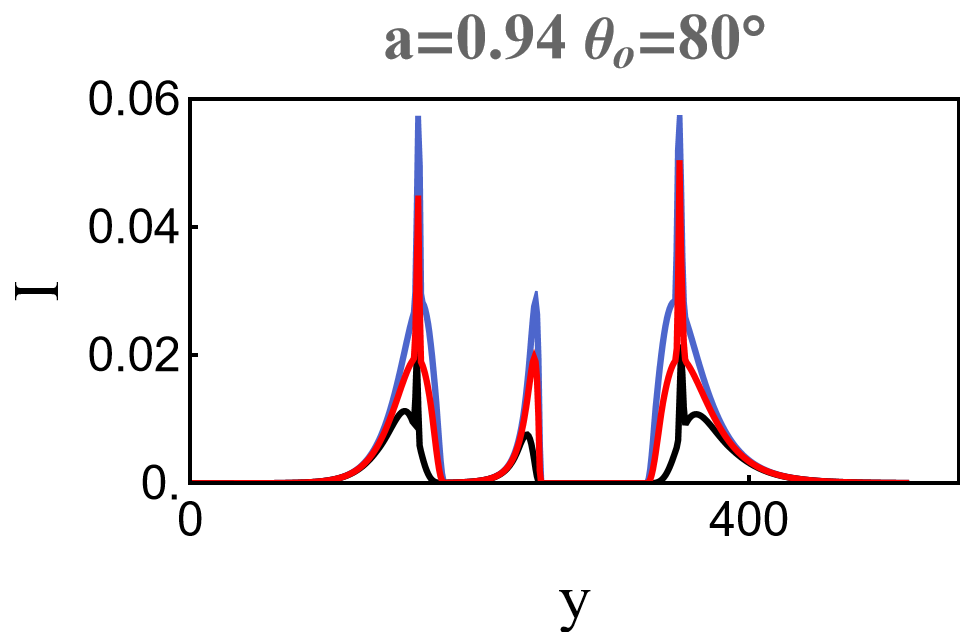}
    };
  \end{tikzpicture}
  \caption{Comparison of the intensity distributions along the $x$- and $y$-axes for different values of $\ell$, where $a=0.94, \ell=-0.99$ (black line), Kerr (red line), $\ell=0.99$ (blue line).}
  \label{fig:11}
\end{figure}

It can be intuitively seen from the Fig.\ref{fig:11} that, as the LV parameter $\ell$ increases, the first and second peak positions along the $x$-axis both shift to the right. The intensity of the first peak gradually increases, whereas that of the second peak decreases. In contrast, along the $y$-axis, the positions of the two peaks remain nearly unchanged, while their peak intensities gradually increase. This indicates that the LV parameter also has a significant influence on the intensity distribution of the thin-disk image.

\begin{figure}[htbp]
    \centering
    \begin{subfigure}[b]{0.32\textwidth}
    \includegraphics[width=\textwidth]{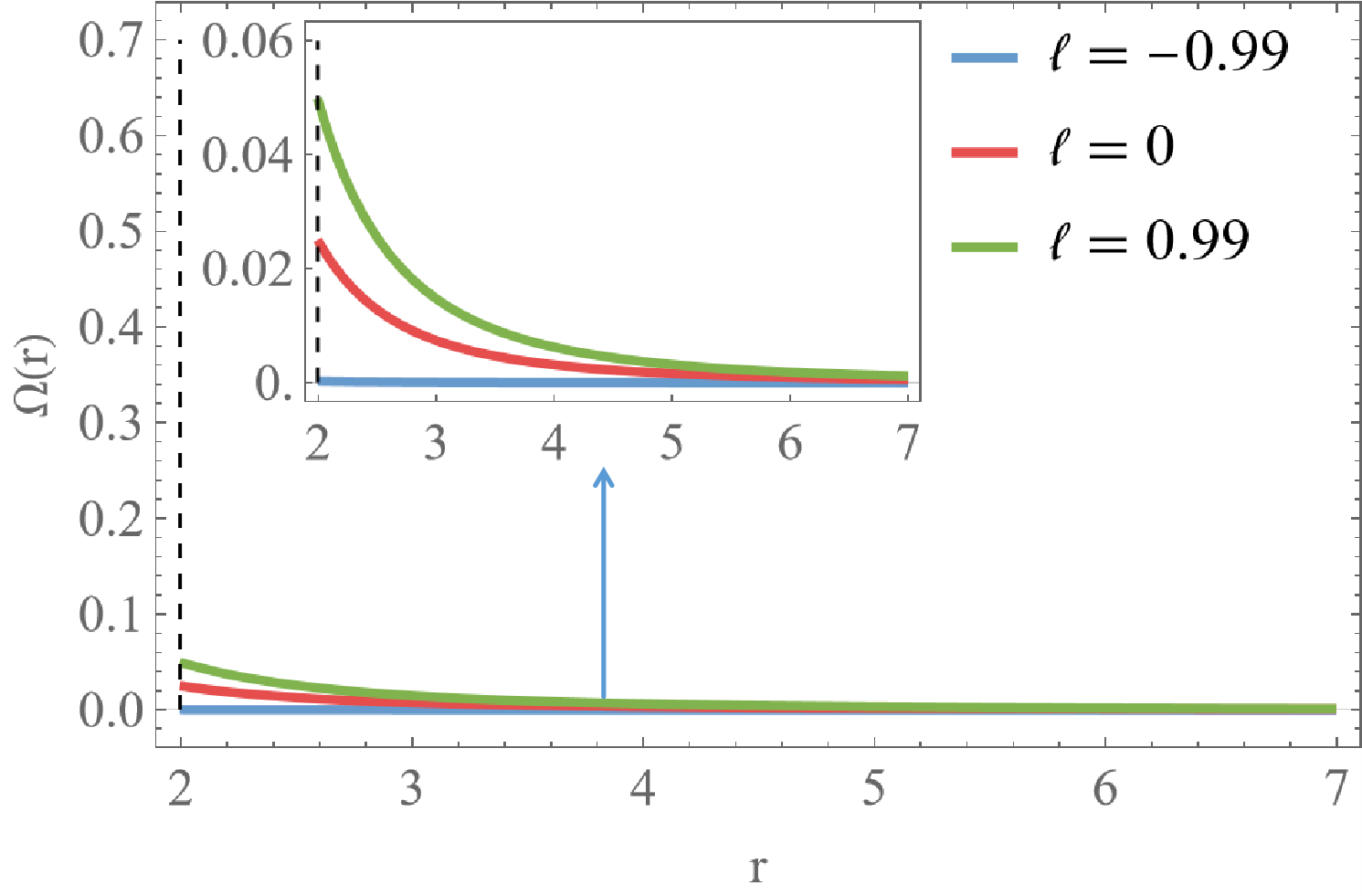}
      \caption{$a=0.1$}
    \end{subfigure}
        \begin{subfigure}[b]{0.32\textwidth}
    \includegraphics[width=\textwidth]{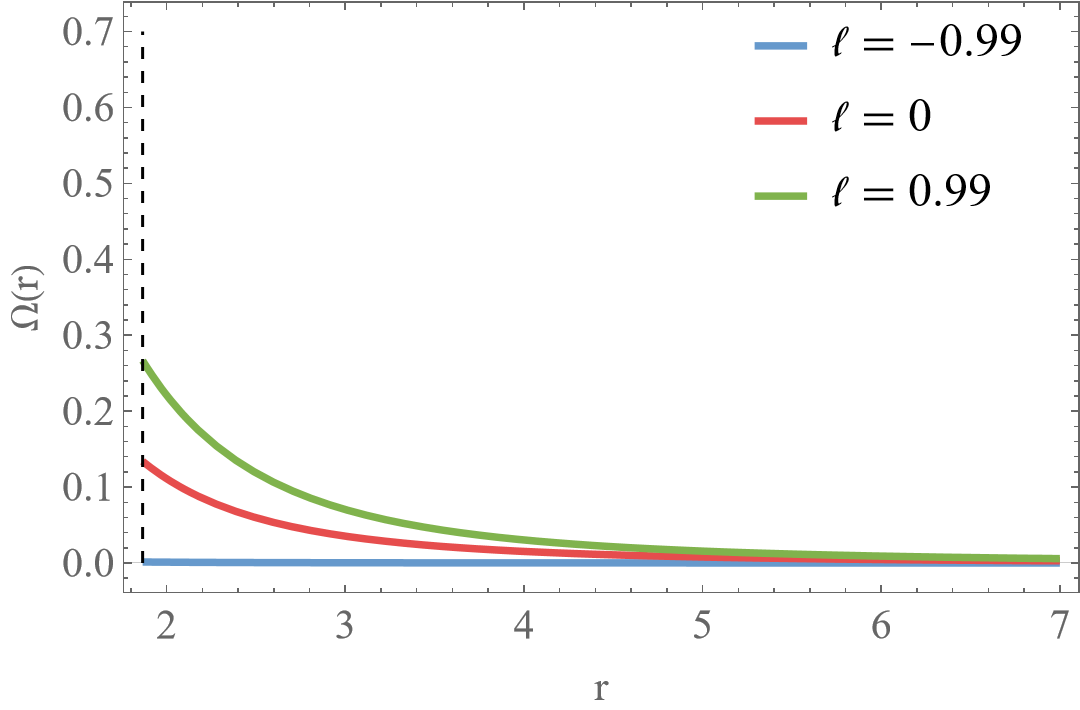}
      \caption{$a=0.5$}
    \end{subfigure}
        \begin{subfigure}[b]{0.32\textwidth}
    \includegraphics[width=\textwidth]{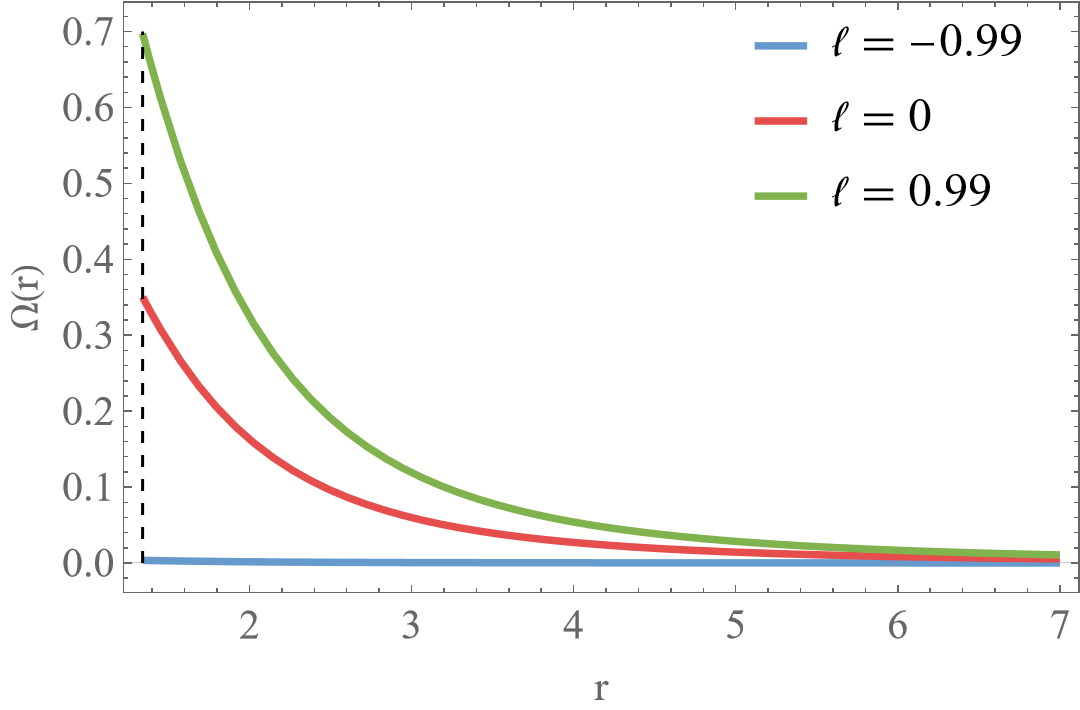}
      \caption{$a=0.94$}
    \end{subfigure}
    \caption{Angular velocity $\Omega(r)$ as a function of $r$ for different values of $\ell$.}
\label{fig:12}
\end{figure}

To further investigate, based on Fig.\ref{fig:10}, it can be observed that when $\ell = 0.99$, the parameter $\ell$ appears to enhance the spin effect of the black hole, whereas for $\ell = -0.99$, it tends to suppress the spin effect. In view of this, we further analysed the angular velocity of the black hole and found that the parameter $\ell$ has a pronounced influence on the angular velocity of the LV black hole. The detailed results are presented in Fig.\ref{fig:12}.
\begin{minipage}[t]{0.32\textwidth}
\end{minipage}

As shown in Fig.\ref{fig:12}, $\Omega(r)$ decreases monotonically with the increase of $r$, indicating that the dragging effect is most pronounced near the event horizon and gradually diminishes to zero at larger radius. When the spin parameter $a$ is small, the influence of $\ell$ on the angular velocity is weak, whereas for $a = 0.94$, the effect becomes significant. More importantly, compared with the Kerr case, we find that for the same radial position, a positive $\ell$ (e.g., $\ell = 0.99$) considerably enhances the angular velocity, while a negative $\ell$ (e.g., $\ell = -0.99$) suppresses it, leading to a very small angular velocity near the event horizon. The curves in the third subfigure correspond to the results shown in Fig.\ref{fig:10}. In summary, the sign of the parameter $\ell$ determines whether the LV effect enhances or suppresses the black hole’s angular velocity. Based on these observable deviations from Kerr black hole, we conclude that the LV black hole can be effectively identified through a combined analysis of its thin-disk image and polarization pattern, providing a potential test for LV effects and the Ho\v{r}ava gravity theory.

\section{Conclusion} \label{sec7}
In this paper, we use the backward ray-tracing method to investigate the image and polarization features of rotating LV black holes within the framework of a thin-disk accretion model. Specifically, within the ZAMO framework, we first employed an optically and geometrically thin accretion disk as the radiation source and numerically solved the geodesic equation of photon. Then, by means of the stereographic projection technique, we simulated the thin-disk images of the black hole. Finally, taking into account the parallel transport of the polarization vector along the geodesics, we obtained the corresponding polarization features of the LV black hole. 

The results show that, when the observed angle $\theta_o = 80^\circ$, the inner shadow appears nearly semicircular, and the critical curve is almost circular for $a = 0.1$. In contrast, for $a = 0.94$, the inner shadow becomes flatter and shifts leftward, while the corresponding critical curve exhibits a distinct ``D”-shaped structure. When $a = 0.94$ and the LV parameter $\ell = 0.99$, the inner shadow becomes more distorted and shifts further to the left compared with the Kerr case. However, when $\ell = -0.99$, the inner shadow evolves in the opposite direction, forming a semi-elliptical shape, and the corresponding critical curve transitions from a ``D” shape to an elliptical one. Meanwhile, as $\ell$ increases from $-0.99$ to $0.99$, the redshifted and blueshifted regions corresponding to the lensing and direct images gradually decrease. Regarding the intensity distribution of the thin-disk image, the peak positions along the $y$-axis remain almost unchanged with varying $\ell$, whereas along the $x$-axis, the peak positions shift rightward as $\ell$ increases, and the peak intensities generally increase, except for the second peak along the $x$-axis.
For the polarization patterns of LV black hole, it is found that, compared with Kerr black hole, when the parameter $\ell = 0.99$, the variation trend of the polarized  intensity is generally consistent with the brightness distribution of the thin-disk image. However, since the parameter $\ell$ affects the intensity distribution of the thin-disk image, it consequently has a direct impact on the distribution of polarized intensity.
Also, the parameter $\ell$ has a significant influence on the polarization direction. 
At the same spatial locations, the polarization direction exhibits noticeable changes, and pronounced deviations are also observed near the critical curves. In the case of $\ell = -0.99$, the polarized intensity and direction still vary significantly with $\ell$. Compared with the case of $\ell = 0.99$, not only does the distribution of polarized  intensity differ, but the polarization direction near the respective critical curves also shows clear distinctions.
On this basis, we further examined the specific influence of the LV parameter $\ell$ on the angular velocity of black hole.
It shows that the value of $\ell$ has a significant impact on the black hole’s angular velocity: when $\ell = 0.99$, the angular velocity is notably enhanced, whereas $\ell = -0.99$ leads to its suppression. Overall, the sign of $\ell$ determines whether the LV effect enhances or suppresses the black hole’s angular velocity.
Combining the above findings, it is true that the LV effect has a significant impact on the observable properties of black hole. It influences the black hole’s angular velocity, which in turn affects the inner shadow and intensity distribution of the thin-disk image, as well as the intensity and direction of the polarized light.

With the continuous upgrades of the EHT, the precision of astronomical observations will be significantly improved, and the resolution of black hole's image will be greatly enhanced. Consequently, we conclude that it will become possible to simultaneously employ the image features of the thin disk and the polarization patterns of black hole to test the LV effect, infer the sign of its parameter, and further examine the validity of the Ho\v{r}ava gravity theory.
It is worth noting that extending the study to the image of black holes in scenarios involving the thick disk, jet, or hotspot is also of great significance, as such investigations may provide additional observational evidence for testing the LV effect.

\section*{Acknowledgments}
Acknowledgments:This work is supported by the National Natural Science Foundation of China (GrantNo.12505059), and by the Sichuan Science and Technology Program (2024NSFSC1999).

\bibliographystyle{utphys}
\bibliography{note}
\end{document}